\documentclass[12pt]{article}
\usepackage{epsfig,amssymb,amsmath,psfrag}

\def \bl  {\begin{align*}}
\def \el  {\end{align*}}

\def \be  {\begin{equation}}
\def \ee  {\end{equation}}
\def \ba  {\begin{eqnarray}}
\def \ea  {\end{eqnarray}}
\def \baa {\begin{eqnarray*}}
\def \eaa {\end{eqnarray*}}
\def \bb  {\begin {thebibliography} }
\def \eb  {\end{thebibliography}}
\def \lab #1 {\label{#1}}

\newcommand{\nn}{\nonumber}
\newcommand{\beq}{\begin{equation}}
\newcommand{\eeq}{\end{equation}}
\newcommand{\beqa}{\begin{eqnarray}}
\newcommand{\eeqa}{\end{eqnarray}}

\def\cN{\mathcal{N}}

\def\cW{\mathcal{W}}

\def\BB{\mathcal{B}}
\def\FF{\mathcal{F}}

\def\MM{\mathcal{M}}
\def\NN{\mathcal{N}}
\def\OO{\mathcal{O}}

\def\l<{\langle}
\def\r>{\rangle}

\def\XXint#1#2#3{{\setbox0=\hbox{$#1{#2#3}{\int}$}
     \vcenter{\hbox{$#2#3$}}\kern-.5\wd0}}

\renewcommand{\title}[1]{\vbox{\center\LARGE{#1}}\vspace{5mm}}
\renewcommand{\author}[1]{\vbox{\center#1}\vspace{5mm}}

\def\Li{ {\rm Li} }

\begin{document}

\thispagestyle{empty}

\begin{flushright}
HU-EP-12/12
\end{flushright}

\vskip2.2truecm
\begin{center}
\vskip 0.2truecm {\Large\bf
{\Large Differential equations for multi-loop integrals \\and two-dimensional kinematics}
}\\
\vskip 1truecm
{\bf
}

\vskip 0.4truecm

\vspace{1.3cm}

\begingroup\bf\large
L. Ferro
\endgroup
\vspace{0mm}

\begingroup
\textit{Institut f\"ur Physik, Humboldt-Universit\"at zu Berlin, \\
Newtonstra{\ss}e 15, D-12489 Berlin, Germany}\par
\texttt{ferro@physik.hu-berlin.de\phantom{\ldots}}
\endgroup

\end{center}

\vskip 1truecm 
\centerline{\bf Abstract} 
In this paper we consider  
multi-loop integrals appearing in MHV scattering amplitudes of planar $\NN = 4$ SYM. 
Through particular differential operators which reduce the loop order by one, we present explicit equations for the two-loop eight-point finite diagrams which relate them to massive hexagons. After the reduction to  two-dimensional kinematics, we solve them using symbol technology. The terms invisible to the symbols are found through boundary conditions coming from double soft limits. These equations are valid at all-loop order for double pentaladders and allow to solve iteratively loop integrals given lower-loop information.
 Comments are made about multi-leg and multi-loop integrals which can appear in this special kinematics.
The main motivation of this investigation is to get a deeper understanding of these tools in this configuration, as well as for their application in general four-dimensional kinematics and  to less supersymmetric theories.

\medskip

 \noindent

\newpage
\setcounter{page}{1}\setcounter{footnote}{0}

\section{Introduction}

Scattering amplitudes in planar $\cN = 4$ super Yang-Mills (SYM) have received a lot of attention in recent years, mostly due to the high amount of symmetry of this theory.  This characteristic leads to many simplifications which give hope to exactly solve the theory.  
It was indeed discovered that the theory possesses a hidden dual conformal symmetry \cite{Drummond:2006rz}-\cite{Drummond:2007au}, not visible at the Lagrangian level, which is related to the duality between maximally helicity-violating (MHV) amplitudes and light-like Wilson loops \cite{Alday:2007hr, Drummond:2007aua,  Drummond:2007cf}, \cite{Drummond:2007au}-\cite{Drummond:2008aq}. This was revealed at weak coupling by considering the amplitudes in a dual coordinate space, defined through the change of variables $p_i^{\alpha\dot\alpha} = x_i^{\alpha\dot\alpha}-x_{i+1}^{\alpha\dot\alpha}$. 
This symmetry extends to the supersymmetric level \cite{Drummond:2008vq, Brandhuber:2008pf} 
 and,
together with the standard superconformal symmetry, forms a Yangian structure \cite{Drummond:2009fd}.
Yangian symmetry constrains and completely fixes the amplitude structure at tree level, when collinear anomalies are also taken into account,
but at loop level it is broken by infrared divergences.
 This was made more explicit by considering the Grassmannian \cite{ArkaniHamed:2009dn,Mason:2009qx}, which gives all possible invariants \cite{Drummond:2010uq,Korchemsky:2010ut}
 under the free Yangian generators.
 At loop level the all-loop planar integrand manifests the full Yangian symmetry \cite{ArkaniHamed:2010kv} but at the level of the actual amplitudes, the invariance is broken
 when integrating over contours.
 Therefore at loop level, even though much progress has been done, finding analytic results for amplitudes seems to be very complicated. 
The multi-parameter nature of the loop integrals increases so enormously their complexity that only very recently an analytical expression for 
the six-point two-loop remainder function was proposed \cite{DelDuca:2009au}. 
Efficient methods to evaluate the integrals appearing in the amplitudes are therefore relevant and can turn out to be very important even for less supersymmetric theories.

In this respect, in \cite{Drummond:2010cz} it was presented a new type of second-order differential equations for on-shell loop integrals. 
The important feature of these equations is their iterative structure, as they reduce the loop degree by one.
In a series of recent papers, the action of differential operators has been proven to be very successful in finding  
the six-dimensional hexagon integrals \cite{Dixon:2011ng,DelDuca:2011wh}, which are related to four-dimensional MHV amplitudes.
Moreover, all-loop differential equations for the BDS-subtracted amplitude have been proposed  \cite{CaronHuot:2011kk}, which give the action of 
superconformal and dual superconformal generators. Even more recently \cite{Paulos:2012nu}, Mellin space representation of amplitudes has been used to derive differential equations for dual conformal integrals.

Meanwhile, a very powerful mathematical tool has been applied to amplitude results, the symbol \cite{Goncharov:symbols} of pure functions. It allowed a remarkable simplification  
\cite{Goncharov:2010jf} of the six-point two-loop remainder function found in \cite{DelDuca:2009au}, as the complicated identities between polylogarithms are trivialized. 
The integrals for scattering amplitudes in $\cN=4$ SYM exhibit indeed an iterated structure and all known perturbative results  can be expressed in terms of generalized 
polylogarithms, multi-dimensional iterated integrals of uniform transcendentality.
Symbols have been efficiently used in the evaluation of the  two-loop $n$-point remainder function \cite{CaronHuot:2011ky} and to make Ans\"atze on the three-loop six-point  remainder function \cite{Dixon:2011pw} and the  two-loop six-point  ratio function \cite{Dixon:2011nj}. The symbol loses nevertheless some information about the function 
 (for some recent development using Hopf algebra of multiple polylogarithms see \cite{Duhr:2012fh}; this method encodes also the $\zeta$ value information). It is indeed invisible to terms which, to a given transcendentality  degree, are lower-degree functions multiplied by constants of the appropriate degree. Therefore other tools must be thought to reconstruct the complete result. We will show the usefulness for this scope of boundary conditions given by soft limits.

Further important achievements have been reached considering a special kinematics, where the external momenta live in two dimensions.
This  kinematics has been used at strong coupling, when considering Wilson loops with 
light-like contours contained in a two-dimensional subspace of Minkowski space-time \cite{Alday:2009yn}. Differential equations in this regime have been proposed in \cite{Alday:2010jz} for two-loop eight-point MHV amplitudes.
The analitycal expressions for two-loop Wilson loops with arbitrary number of points 
\cite{Heslop:2010kq} and three-loop eight-point Wilson loops \cite{Heslop:2011hv} have been evaluated.
At least in the one-loop and two-loop cases only logarithms enter in the Wilson loop expressions   \cite{Alday:2009yn,DelDuca:2010zp,Heslop:2010kq}.
In this kinematics it has also been possible to demonstrate the Yangian invariance of Wilson loops at one-loop  \cite{Drummond:2010zv}.

The main motivation for this paper is to combine  these new tools in the search for efficient and powerful methods to find analytic results for multi-loop multi-leg diagrams appearing in scattering amplitudes.
We will focus our attention on MHV scattering amplitudes in the basis given in  \cite{ArkaniHamed:2010kv, ArkaniHamed:2010gh}. 
Starting with the two-loop eight-point finite diagrams, we will present the second-order differential operators, in momentum twistor variables \cite{Hodges:2009hk},
 which lower the loop degree by one relating them to hexagon integrals. Then we will truncate the differential equations to the restricted two-dimensional kinematical regime. 
This configuration simplifies the results obtained and allows to analitically solve the equations, when boundary conditions are taken into account.  The same equations can be used for higher loops, extending the results of \cite{Alday:2010jz}.
Through these results, beyond the explicit analytic expressions of the diagrams, we also want to give further contribution to the investigation of these methods. First of all, it is becoming more and more challenging to understand which letters can enter in a symbol subject to physical constraints, such as \emph{e.g.} the operator product expansion \cite{Alday:2010ku}-\cite{Sever:2011pc}. In this respect,  more data are needed to possibly constraining them. Another important goal is the analysis of boundary conditions, coming for instance from soft limits as we will discuss. The differential equations and the symbols leave indeed some ambiguities about the function, which must be fixed demanding additional constraints. 
The differential equations discussed here, as already stressed, are valid for an infinite series of loop integrals which  are  part of higher-loop MHV  amplitudes. Behind them, other topologies appear, as well as in NMHV amplitudes \cite{ArkaniHamed:2010gh}, where also different tensor numerators show up. New differential operator are therefore necessary and further investigation needed.

The paper is organized as follows. The first part is an introduction on the subject, in order to review the main ingredients  which will be used throughout the paper. 
In particular we recall the dual conformal integrands appearing in one-loop and two-loop MHV amplitudes and how specific differential operators act on their master topology. We also review the two-dimensional kinematics. We then discuss the boundary conditions that can be used to fix the lower transcendental-degree functions.  Then in section \ref{Fc}, an explicit example is presented, appearing in the  two-loop eight-point MHV amplitude. We describe the action of a particular operator in momentum twistor space and then reduce it to the two-dimensional kinematics in order to find an explicit result.  In  section \ref{othertop} we summarize the results  for the other finite diagrams appearing in the two-loop eight-point case. 
 We then discuss the generalization of such tools for the cases of two-loop $n$-point integrals and $L$-loop double pentaladder integrals.
In the appendices we review the notion of symbols and give technical details.

\section{Background material and setup}
\label{intro}

We recall here some basics about MHV scattering amplitudes in momentum twistor space, differential operators for on-shell integrals and two-dimensional kinematics. Later we will discuss boundary conditions which are helpful in evaluating diagrams from differential equations and symbols.
As we already pointed out in the introduction, many developments have been reached in studying MHV amplitudes. By using a BCFW-generalized recursive formula for the all-loop integrand, it has been possible to reconstruct the integrand of multi-loop amplitudes  \cite{ArkaniHamed:2010kv,ArkaniHamed:2010gh}. 
In the case of one- and two-loop MHV amplitudes, a single master topology contributes, namely the tensor-pentagon diagram.  
The one-loop case is depicted in (\ref{basis1}), which is "formally" dual conformal invariant, where "formally" means that IR divergences can appear when the integrals are performed. The dashed line represents the tensor numerator and its explicit expression will be discussed later on. The box-integrals are "boundary terms" of this topology, which are retrieved by taking soft limits \cite{Drummond:2010cz}.
\beq
\label{basis1}
A_{\mathrm{MHV}}^{\mathrm{1-loop}} = \sum_{j<k}\qquad
{
\parbox[c]{30mm}{\includegraphics[height = 4cm] {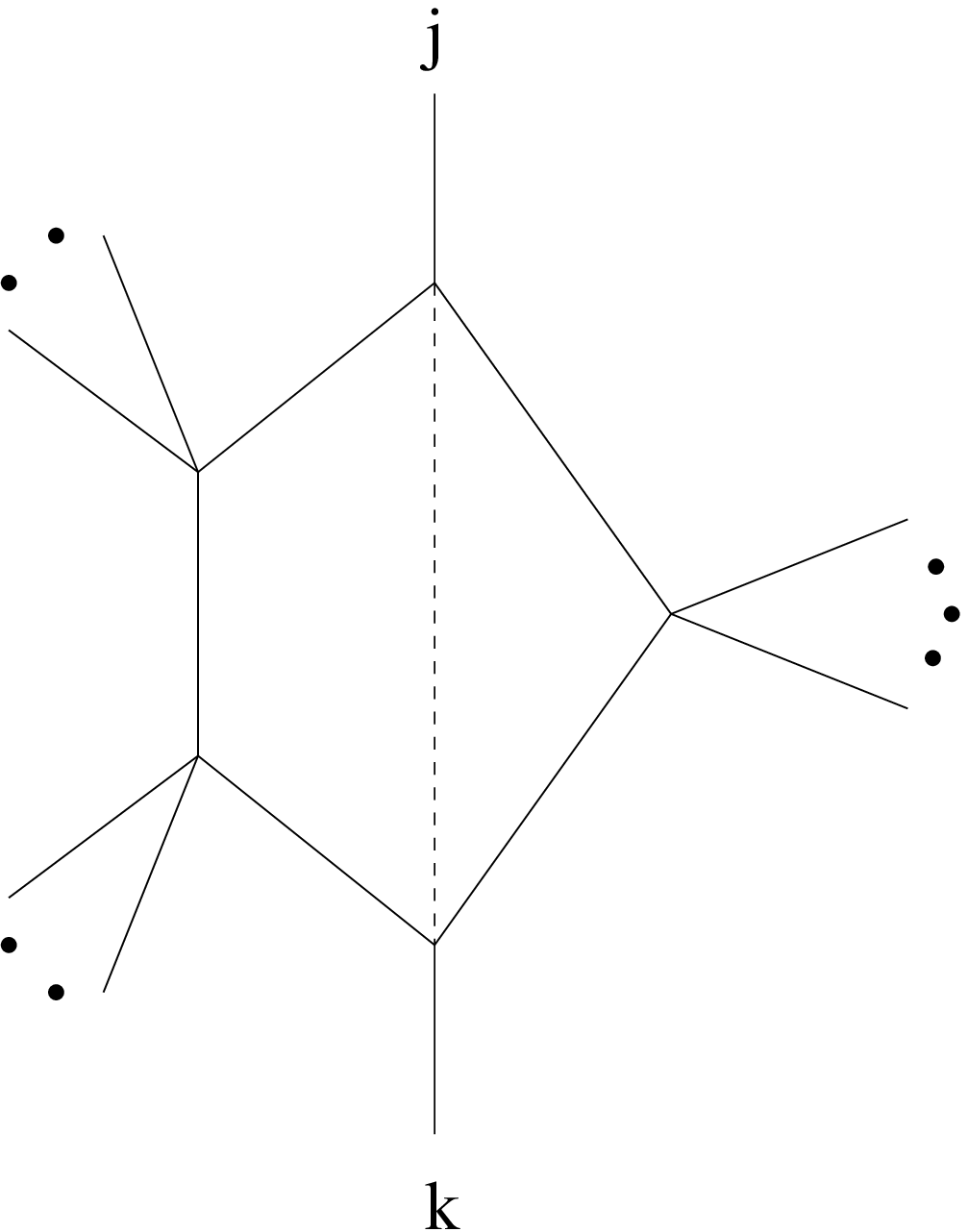}}
}
\eeq
The same topology appears at two-loops (\ref{basis2}), where the tensor pentagon is now a subintegral and penta-boxes, as well as double-boxes are again boundary terms:
\beq
\label{basis2}
A_{\mathrm{MHV}}^{\mathrm{2-loop}} = \frac{1}{2} \sum_{i<j<k<l}\qquad
{
\parbox[c]{30mm}{\includegraphics[height = 4cm] {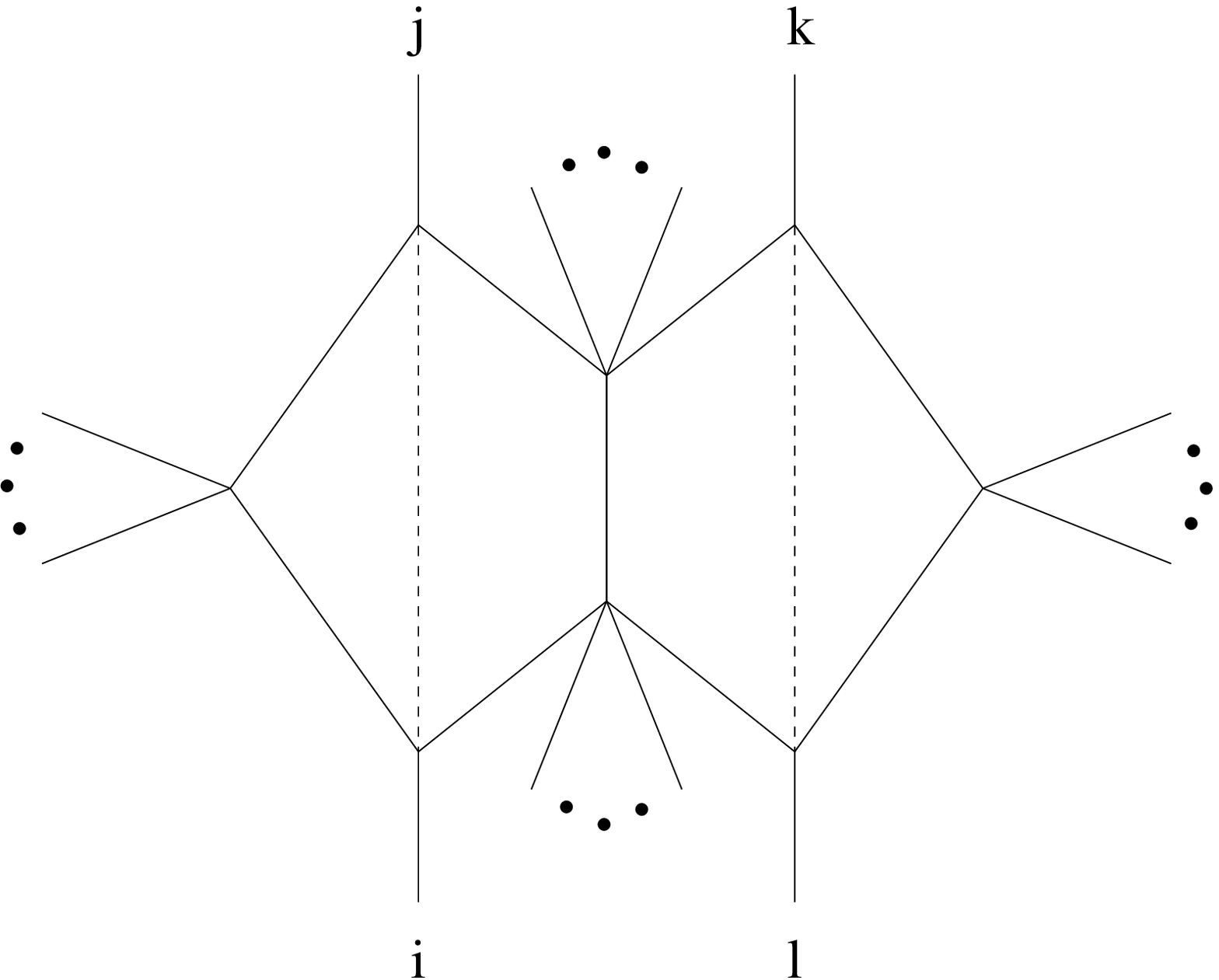}}
}~
\eeq

There are different spaces where amplitudes can be defined in and  in the following we want to review them briefly, in order to show the relation of momentum twistor variables to the dual space. In the spinor helicity formalism, the massless momenta are written as $p_i^{\alpha\dot\alpha} = \lambda_i^{\alpha}\tilde\lambda_i^{\dot\alpha}$, where $\lambda_i^{\alpha}$ and $\tilde\lambda_i^{\dot\alpha}$ are two-component commuting spinors and define the on-shell space. The dual coordinates $x_i$ are related to the momenta through 
\beq
p_i^{\alpha\dot\alpha} = x_i^{\alpha\dot\alpha}-x_{i+1}^{\alpha\dot\alpha} \, ,
\eeq  
and describe a light-like closed polygon.
The momentum twistor variables  \cite{Hodges:2009hk} are the twistors associated to the dual coordinates $x_i$'s and are defined as follows:
\be
Z_i^A = (\lambda_i^\alpha , \mu_i^{\dot\alpha}), \qquad \mu_i^{\dot\alpha} = x_i^{\alpha \dot\alpha} \lambda_{i \alpha} = x_{i+1}^{\alpha \dot\alpha} \lambda_{i \alpha}\,
\ee
where the second relation is the so-called incidence relation and the index $A$ is in the fundamental of $sl(4)$. The most powerful property of momentum twistors is that they are free variables, since momentum conservation  and light-likeness constraint are already encoded in the definition.
The dual distances are therefore expressed in the following form
\be
x_{ij}^2 = \frac{(i-1\,\, i\,\, j-1\,\, j)}{\langle i-1 \, i \rangle \langle j-1 \, j\rangle}\,
\label{dualTotw}
\ee
where the four-bracket
\be
(ijkl)  = \epsilon_{ABCD} Z_i^A Z_j^B Z_k^C Z_l^D
\ee
is dual conformal invariant, while $\langle i \, j \rangle = \epsilon_{\alpha\beta}\lambda_i^{\alpha}\lambda_j^{\beta}$ is invariant under Lorentz and (dual) translations only. To a point $x_i$ corresponds therefore  the line $(Z_{i-1} Z_i)$, while two light-like separated dual points are mapped to intersecting lines in twistor space. 

\begin{figure}[t]
\psfrag{dots}[cc][cc]{$\ldots$}
 \centerline{
 {\epsfxsize3cm  \epsfbox{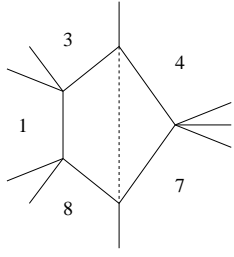}}
} 
 \caption{\small One-loop tensor pentagon integral $F_{9}^{(1)}$. Dual points $i\equiv x_i$'s are showed.} \label{fig-massivepent}
\end{figure}

Let us now consider a particular diagram, present in (\ref{basis1}), in momentum twistor space, namely a nine-point one-loop tensor pentagon, see Fig. (\ref{fig-massivepent}) where dual points are indicated. We choose this specific integral to show differential operators at work in an explicit example which will be useful later on. Its expression is
\beq
F_9  ^{(1)} = \int \frac{\mathrm{d}^4 Z_{AB}}{i \pi^2} \frac{N~ (AB37)}{(AB23)(AB34)(AB67)(AB78)(AB91)} \equiv  \int \frac{\mathrm{d}^4 Z_{AB}}{i \pi^2} I_9^{(1)}
\eeq
where $N$ is the normalization factor $N = (1234) (6789)$ and the tensor numerator $(AB37)$ is represented pictorially with a dashed line. All these characteristics assure that this is a pure integral with unit leading singularity \cite{ArkaniHamed:2010kv,ArkaniHamed:2010gh}. The integration is performed over the space of lines in momentum twistor space
\beq
d^4 Z_{AB} \sim \langle AB \rangle^4 d^4x_0 \,.
\eeq
We want now to act on this integral with one of the operators considered in  \cite{Drummond:2010cz} and whose iterative structure allows to relate $L$-loop integrals to lower-loop ones. The twistor derivative is defined to be
\beq
\OO_{ij} = Z_i \cdot \frac{\partial}{\partial Z_j} \, ,
\eeq
which on  four-brackets  simply gives $\OO_{ij} (j l k m) = (i l k m)$. Considering the explicit action of the operator \emph{e.g.} $\OO_{24} \OO_{86} \equiv Z_2 \cdot \frac{\partial}{\partial Z_4} Z_8 \cdot \frac{\partial}{\partial Z_6} $ on the integrand $I_9^{(1)}$ we find
\beq
\OO_{24} \OO_{86}  I_9^{(1)} = \frac{N~ (AB37)}{(AB34)^2 (AB67)^2 (AB91)} \, ,
\eeq
where the normalization factor $N$ has not been touched by the particular operator. Integration using Feynman parametrization gives
\beq
\OO_{24} \OO_{86}  F_9  ^{(1)} \propto \frac{N~ (9137)}{(9134) (9167) (3467)} .
\eeq
This substitution can be used whenever the pentagon $F_9  ^{(1)}$ is a subintegral and pictorially it can be depicted as
\beq
\label{reduction}
\OO_{24} \OO_{86}\qquad
{
\parbox[c]{30mm}{\includegraphics[height = 3cm] {massivepent.eps}}
}
\qquad \propto \qquad
{
\parbox[c]{30mm}{\includegraphics[height = 3cm] {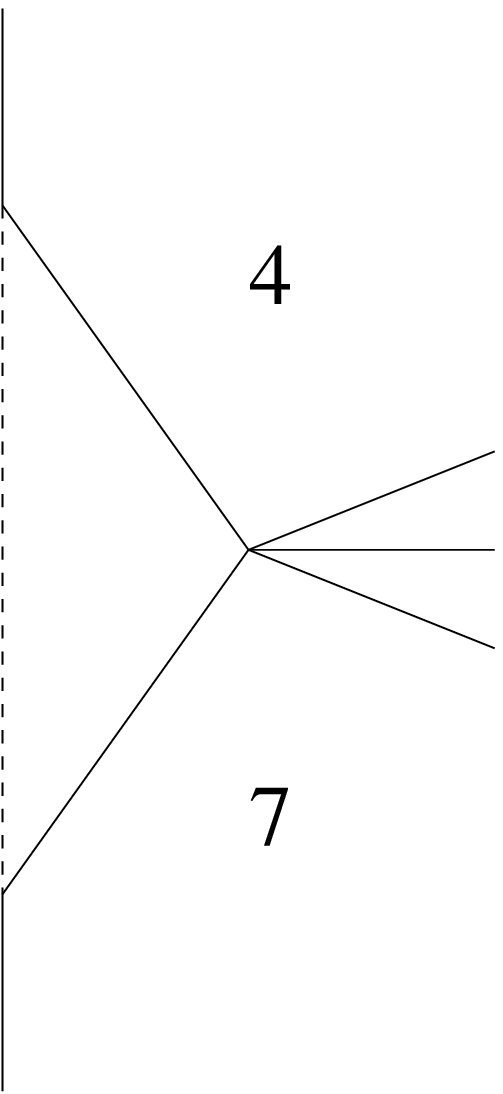}}
}
\eeq
This method is very powerful since the second-order differential operator $\OO_{24} \OO_{86}$ reduces the loop order by one: in the next section we will apply this and other twistor derivatives of the same kind to derive differential equations.
Specifically, we will restrict our analysis to finite integrals; IR divergent diagrams need separate consideration \cite{Drummond:2010cz} and the introduction of \emph{e.g.} the AdS mass regulator of \cite{Alday:2009zm}. 
Since these integrals are dual conformal, they have a restricted variable dependence. Indeed, they can be expressed in terms of the $(3n-15)$ conformal invariant cross-ratios only (in case of three-point functions the integrals are rational functions),
\beq
\label{crossR}
u_{ij} = \frac{x_{i,j+1}^2 x_{i+1,j}^2}{x_{i,j}^2 x_{i+1,j+1}^2} = \frac{(i-1\,\, i\,\, j\,\, j+1) (i\,\, i+1\,\, j-1\,\, j)}{(i-1\,\, i\,\, j-1\,\, j) (i\,\, i+1\,\, j\,\, j+1)} \, ,
\eeq
where (\ref{dualTotw}) is used to translate from dual coordinate space to momentum twistor variables.
One can therefore rewrite the twistor operators in terms of derivatives acting on cross-ratios, by using the chain rule.

We want to end this section by reviewing the two-dimensional kinematics, where the momenta of the external particles lie in a two-dimensional plane. The momentum twistor variables are restricted and preserve two commuting copies of $sl(2)$ inside $sl(4)$. We choose (see appendix \ref{appendix-2dim})
\begin{eqnarray}
Z_{2i+1}=\left( \begin{array}{c}  0 \\  0 \\  Z_{2i+1}^3  \\   Z_{2i+1}^4 \\  \end{array}  \right),~~~
Z_{2i}=\left(  \begin{array}{c}   Z_{2i}^1 \\  Z_{2i}^2 \\    0 \\  0 \\    \end{array}   \right) \, .
\end{eqnarray}
Four-brackets with an odd number of odd or even twistors vanish and many cross-ratios become trivial, in particular $u_{i,i + \mathrm{odd}} = 1$. 
Specifically the number of independent cross-ratios is reduced to $(n-6)$ out of $(3n-15)$ of the original four-dimensional space.

\subsection{Boundary conditions from soft limits }
\label{boundary}
 \begin{figure}[t]
\psfrag{dots}[cc][cc]{$\ldots$}
 \centerline{
 {\epsfxsize3cm  \epsfbox{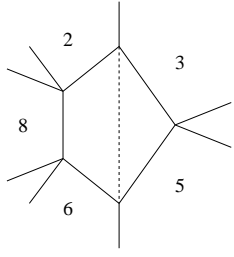}}
} 
 \caption{\small One-loop tensor pentagon integral $\tilde\Psi ^{(1)} (u,v,w)$.} \label{massivepent7}
\end{figure}
In this section we explain  boundary conditions deriving from soft limits, which will be necessary to fix the ambiguities in symbol technology for lower-transcendental functions.
We first consider the eight-point tensor pentagon $\tilde\Psi ^{(1)} (u,v,w)$ of Fig. \ref{massivepent7}
which depends on the following three cross-ratios:
\beq
u = \frac{x^2_{28} x^2_{36}}{x^2_{26} x^2_{38}} \, , \qquad v = \frac{x^2_{68} x^2_{25}}{x^2_{26} x^2_{58}} \, ,  \qquad w = \frac{x^2_{35} x^2_{26}}{x^2_{36} x^2_{25}} \, ,
\eeq
and whose expression is known \cite{Drummond:2010cz}:
\beq
\tilde\Psi ^{(1)} (u,v,w) = \log u \log v + \Li_2(1-u) + \Li_2(1-v) + \Li_2(1-w) -  \Li_2(1-u w) - \Li_2(1-v w) \, .
\eeq
A simple soft limit consists on taking $p_4 \rightarrow 0 $ ($w\rightarrow0$) to relate the eight-point pentagon integral with the seven-point one.
Another  limit, which will be useful later on, is $p_5 \rightarrow 0$. In dual coordinates this means $x_5 = x_6$ and the cross-ratios take the values $(v \rightarrow 1, w \rightarrow 1)$, with $u$ arbitrary. On the function the limit becomes
\beq
\lim_{\tau\rightarrow 0} \tilde\Psi ^{(1)} (u, 1- \tau \xi, 1- \tau) = 0 \, ,
\eeq
where the ratio $\frac{1-v}{1-w} = \xi$ can be arbitrary. This limit, as well as the previous one, can be used also for diagrams where this pentagon is a subintegral.
For instance, the seven-point pentagon case can be used in the six-point double-pentagon diagram $\Omega^{(2)}(v_1, v_2, v_3)$,  defined in \cite{Drummond:2010cz}, to obtain other boundary conditions.
If we now consider the integral $F_b^{(2)}$, see Fig. \ref{fig-twoloop}, appearing in the two-loop eight-point MHV scattering amplitude, we can take the double soft limit $p_5, p_6 \rightarrow 0$. Since we are most interested in the two-dimensional kinematics,  we can use the fixing of appendix \ref{appendix-2dim} to find that the relevant cross-ratios are
\beq
u_1 = \frac{x^2_{27} x^2_{36}}{x^2_{26} x^2_{37}} = \frac{x}{1+x} \, , \qquad  u_2 = \frac{x^2_{38} x^2_{47}}{x^2_{37} x^2_{48}} = \frac{y}{1+y} \, .
\eeq
The double soft limit under consideration implies that  
\beq
(u_1 \rightarrow 1 \, , \, u_2 \rightarrow 0) \, \equiv \, (x \rightarrow \infty \, ,\, y \rightarrow 0) \, .
\eeq
Since the product $(x \, y)$ can be arbitrary,  calling  $x=\frac{1}{\tau}$ and $y = \tau\xi$ we expect
\beq
\label{limit}
\lim_{\tau \rightarrow 0} F_{b}^{(2)}\left(\frac{1}{\tau}, \tau \xi \right) = 0 \, .
\eeq
Due to the symmetry of the diagram, the limit $(x \leftrightarrow y)$ is also valid.
The same mechanisms work for the case $ F_{c}^{(2)}$, which vanish when $(p_5, p_6 \rightarrow 0)$, so that $(x \rightarrow 0 \, , \, y \rightarrow \infty)$, giving a boundary condition similar to (\ref{limit}).
Considering instead the double soft limit $(p_6, p_7 \rightarrow 0)$, this diagram can be related to $\Omega^{(2)}(1,1,1)$.

Appropriate double soft limits give therefore powerful constraints for diagrams appearing in scattering amplitudes and, as we will show, allow to reconstruct the complete result for an integral from its symbol.

\section{Loop amplitudes and differential equations}

 \begin{figure}[t]
\psfrag{dots}[cc][cc]{$\ldots$}
 \centerline{
 {\epsfxsize4cm  \epsfbox{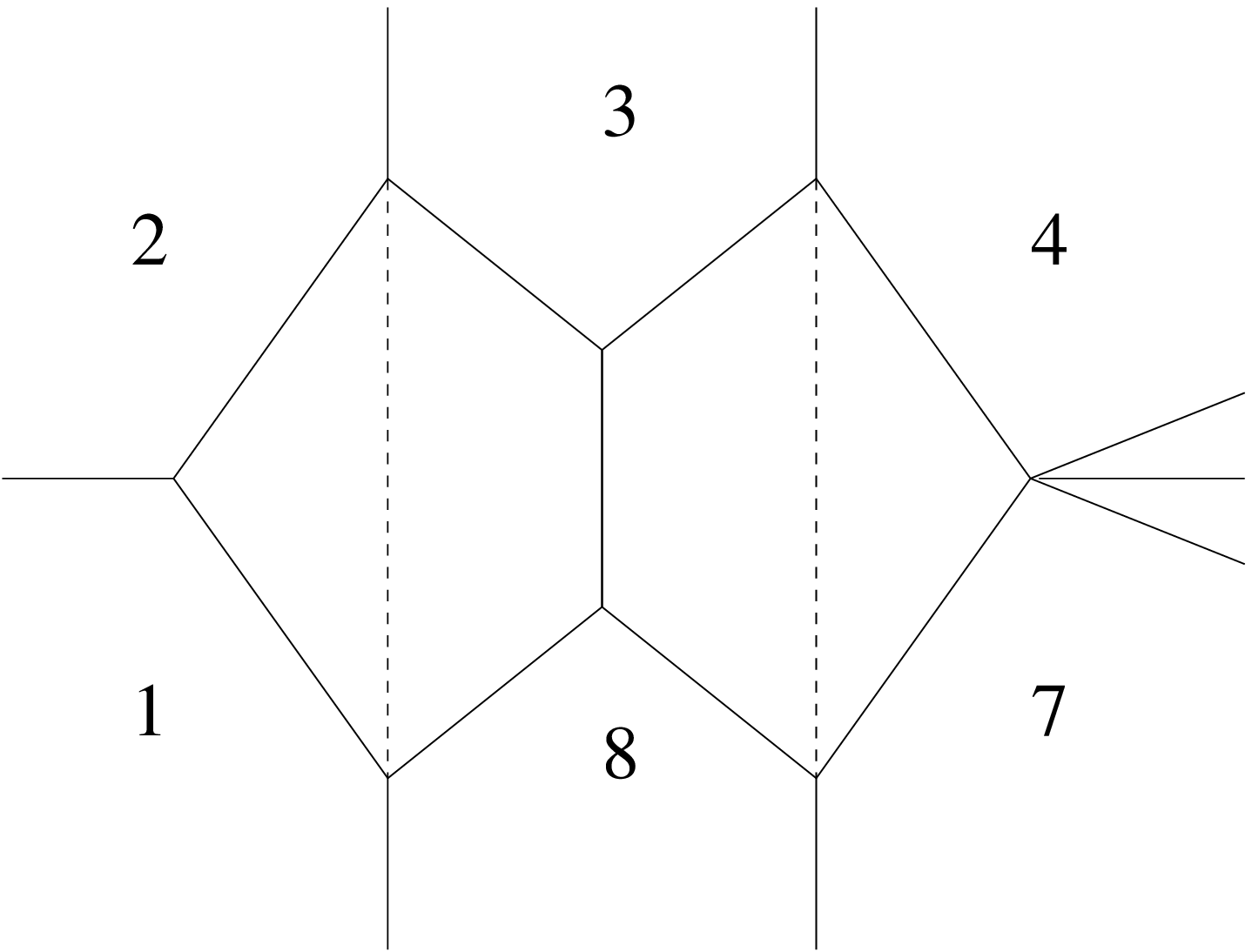}}
\qquad\quad  {\epsfxsize4cm  \epsfbox{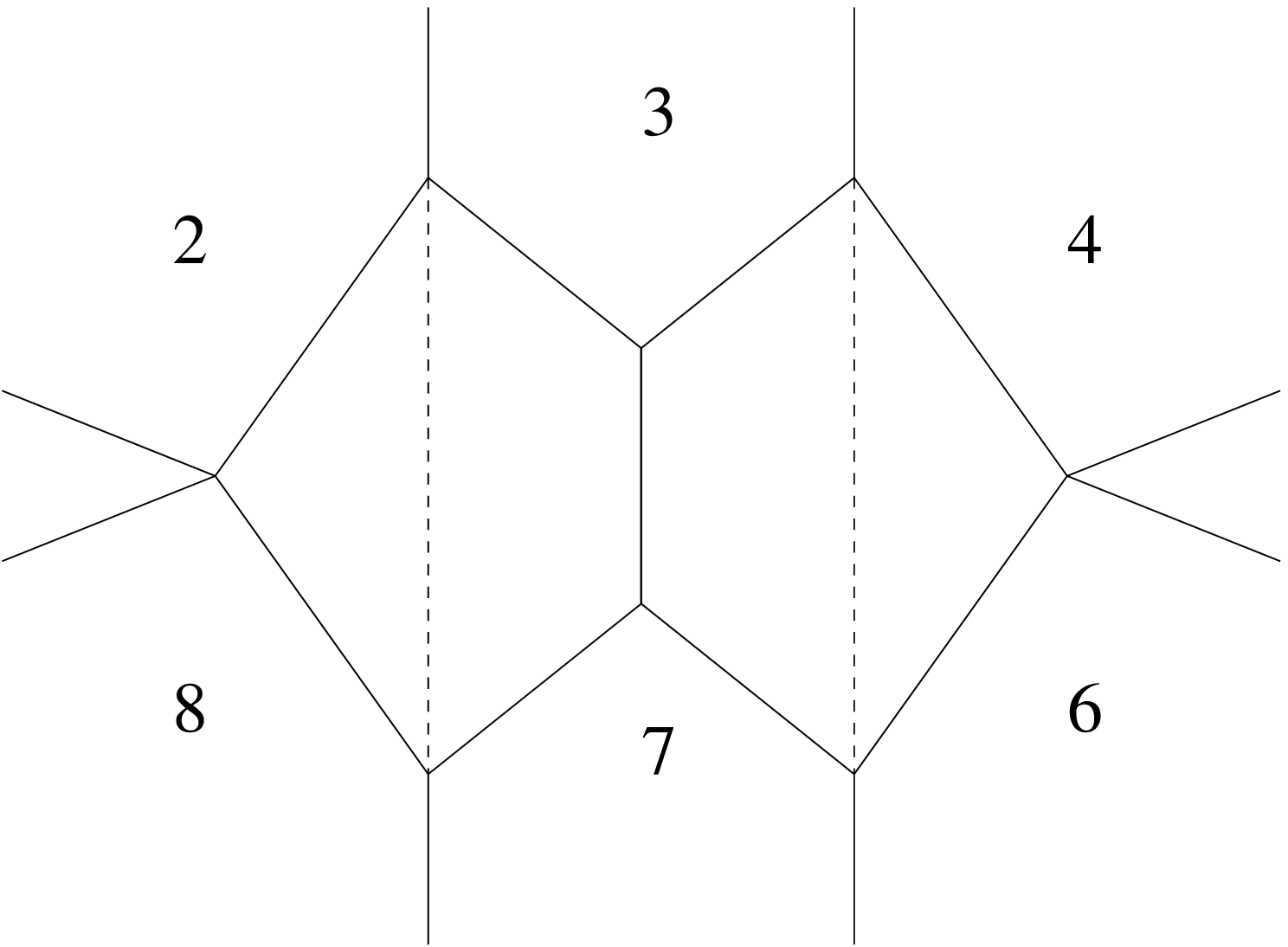}}
\qquad\quad  {\epsfxsize4cm  \epsfbox{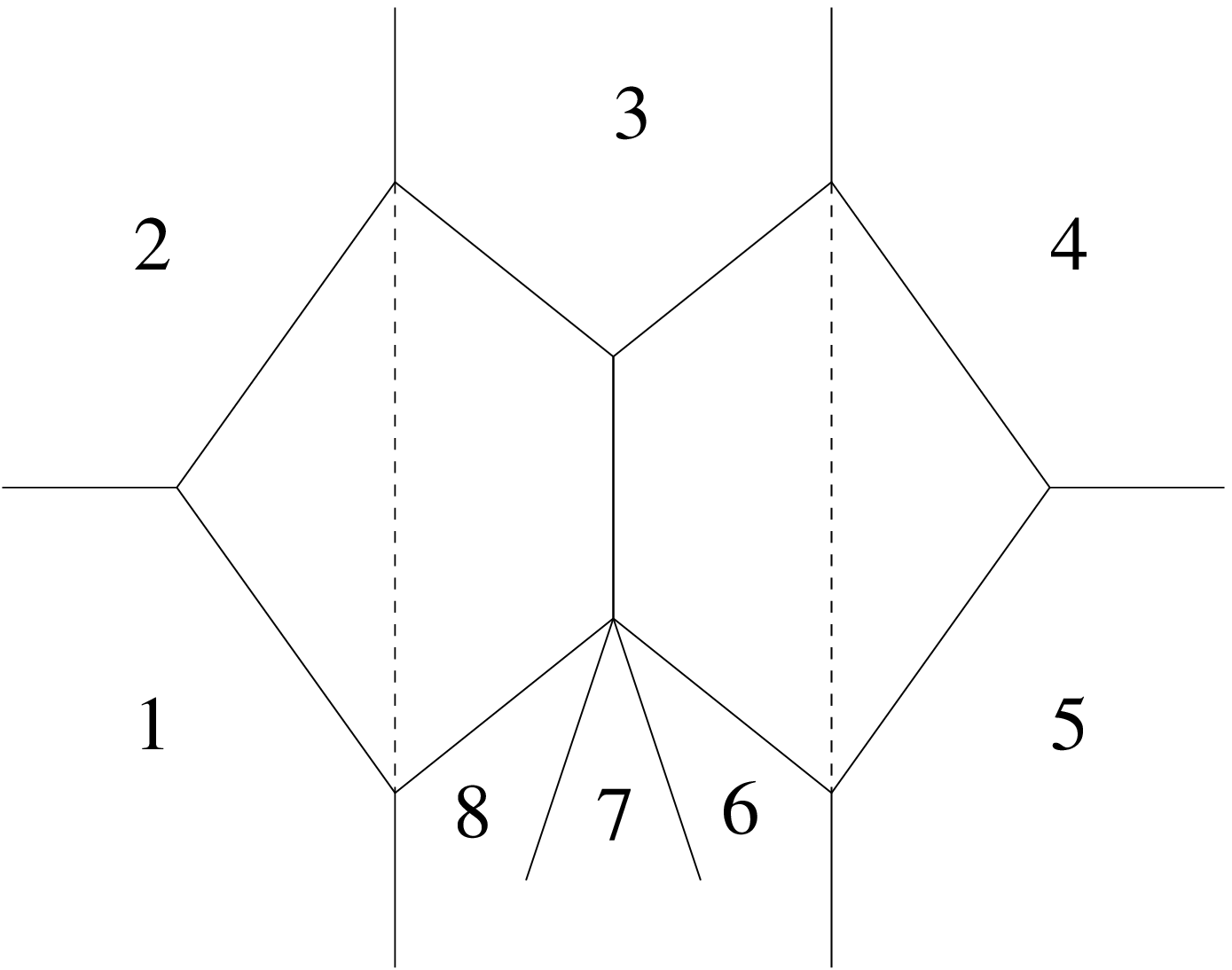}}
} 
 \caption{\small Two-loop integrals  $F_{a,b,c}^{(2)}$ considered in the text. Numbers denote the dual coordinates.
The explicit definitions are given in appendix \ref{defloops}.} \label{fig-twoloop}
\end{figure}

 \begin{figure}[t]
\psfrag{dots}[cc][cc]{$\ldots$}
 \centerline{
 {\epsfxsize2.5cm  \epsfbox{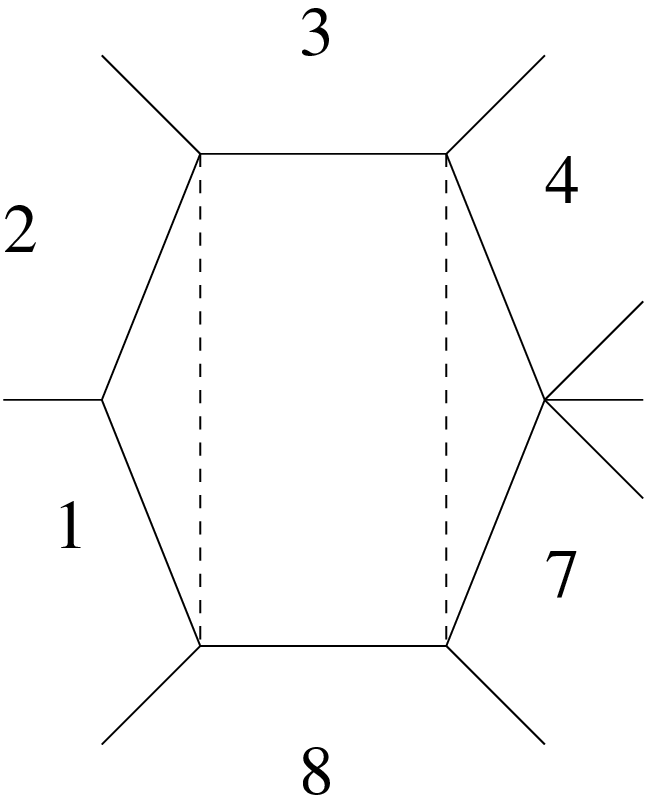}}
  \qquad\quad{\epsfxsize2.5cm  \epsfbox{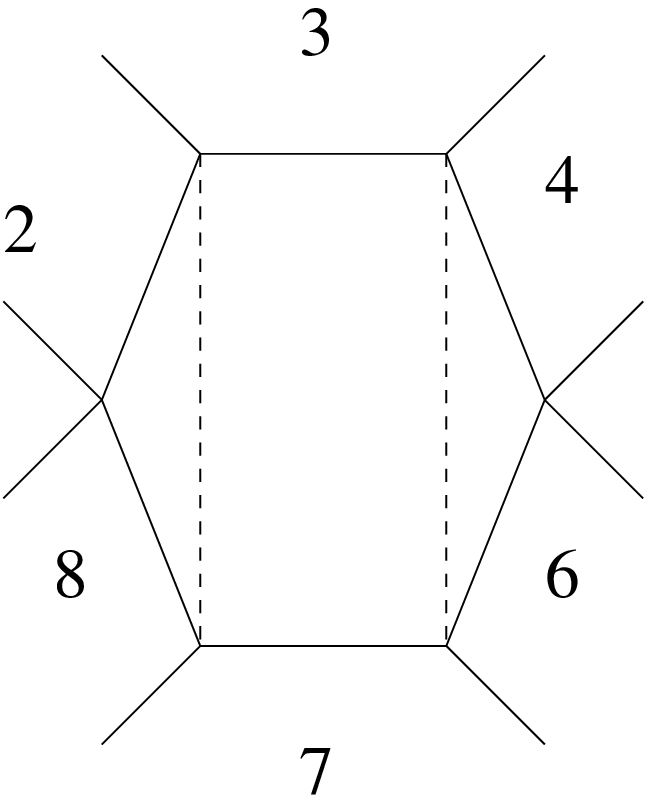}}
  \qquad\quad {\epsfxsize2.5cm  \epsfbox{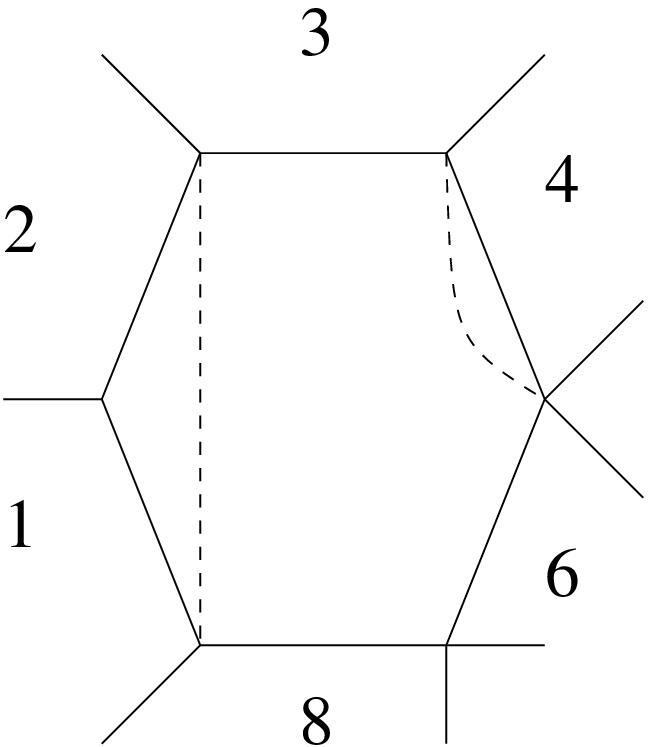}}
}
\caption{\small One-loop hexagons $F_{a,b,c}^{(1)}$ considered in the text. The explicit definitions are given in appendix \ref{defloops}.} \label{fig-oneloop}
\end{figure}

Let us start by considering the finite diagrams which appear in the two-loop eight-point MHV amplitudes and given by the compact expression (\ref{basis2}). Specifically, there are three topologies contributing to the amplitude, which are collected in Fig. \ref{fig-twoloop}. As already mentioned, they are composed by tensor-pentagon diagrams.
Some differential equations for these diagrams have been already found in \cite{Alday:2010jz}; here we want to extend those results in such a way to be possible to use them for higher-loop higher-point diagrams.
We reviewed in the previous section that the number of independent cross-ratios is $(3n-15)$; since the integrals do not depend on some dual points,  the respective cross-ratios are not present and their  number is actually reduced diagram by diagram.
The goal is to find differential operators for each of these diagrams such that  the loop-level is lowered by one: the operators will be therefore valid for the $L$-loop double pentaladder diagrams. We will show that the iterative structure relates them to one-loop hexagon integrals. 

\subsection{Double-pentagon integral $F^{(2)}_{c}$}
\label{Fc}
We will present the $F^{(2)}_{c}$ case in detail here, and resume the results for the others in the next subsection.
 Since this diagram is independent of the dual point $x_7$, the following cross-ratios will be present only:
\begin{eqnarray}
 u_{1}&:=& u_{14} \,, \qquad u_{2} :=  u_{25}  \,, \qquad u_{3} := u_{58}  \, , \nonumber \\
 u_{4}&:=& u_{83} \,, \qquad u_{5} :=  u_{15} \,,\qquad  u_{6} :=  u_{48} \, ,
\end{eqnarray}
where they are defined as in (\ref{crossR}). Going to two-dimensional kinematics by using the fixing (\ref{fixing}), they become
\begin{eqnarray}
\label{kin-2d-c}
u_{1} = u_{2} = u_{3} = u_{4} = 1  \,,\qquad u_{5} = \frac{1}{1+y} \,,\qquad u_{6} = \frac{1}{1+x} \,.
\end{eqnarray}
Since there are two legs attached to one of the gluing points, we can use the following equation \cite{Drummond:2010cz},
\begin{eqnarray}
\label{slingshot-3}
\OO_{24} \OO_{46} \frac{1}{N_{\rm p}} F^{(1)}_{7} =  \frac{(7135)}{(7134)(7156)(3456)} \, ,
\end{eqnarray}
with $ F^{(1)}_{7} $ being the seven-point pentagon integral, \emph{i.e.}  $F_9  ^{(1)}$ with the right-most corner massless.
Note that this is slightly different from the operator taken in account in the previous section, as the second twistor derivative does not commute with the normalization factor $N_{\rm p}$.
We therefore consider the differential equation
\begin{equation}
\OO_{24} \OO_{46} \frac{1}{N_{2c}} F_{c}^{(2)} = \frac{1}{(3456)} \frac{1}{N_{1c}} F_c^{(1)}\,,
\end{equation}
where $N_{2c} = -(1234)\left[(3781)(5624)+(7812)(3456)\right]$ and $N_{1c} = (1234) (6781)$ are the respective normalizations (see appendix \ref{defloops}). $F_c^{(1)}$ is the one-loop hexagon integral with two massive corners as shown in Fig. \ref{fig-oneloop}(c). Therefore the second-order differential operators has related a two-loop double-pentagon diagram with a simpler one-loop hexagon one.
By using the chain rule we obtain
\begin{equation}
\OO_{46} = -\frac{(8145)}{(8156)} [u_5 u_2 (1-u_2)\partial_2 + (1-u_3)\partial_3 + u_5 (1-u_5)\partial_5] \,.
\end{equation}
The expression for the operator $\OO_{24}$ involves four-brackets that are related to the cross-ratios
via quadratic equations. We do not need to present it here explicitly since the coefficients of the derivatives $\partial_1,\partial_3 $ and $\partial_4$ vanish for two-dimensional kinematics, such that
both $\OO_{46}$ and $\OO_{24}$ are compatible with (\ref{kin-2d-c}).
Implementing the choice of momentum twistors as in (\ref{fixing}), the differential equation becomes
\begin{equation}
\label{FcDiff}
 y \partial_{y} (1 + y) \partial_{y}   F_{c}^{(2)}(x,y) =    F_{c}^{(1)}(x,y)  \, .
\end{equation}
The one-loop integral $F_{c}^{(1)}(x,y)$ is calculated from its Feynman parameterization,
\begin{eqnarray} \label{F1c}
 F_{c}^{(1)}(x,y)  &=& \frac{y}{(x - y)} \left[ -\log(y) \log (1+x) + \log (1+y) \log(1+ x)  \right]  \nonumber \\
&+& \frac{x}{(x - y)} \Bigg[ \log(y) \log (1+y) + \log (1+y) \log(x) -\log(1+x) \log(1+y) \nonumber\\
&& \qquad \qquad -\log(x) \log(1+x) + 2 \, \Li_{2}(-y) -2 \, \Li_{2}(-x) \Bigg] \,.
\end{eqnarray}
As discussed before, the boundary conditions for this case are given by the double soft limit $p_5,p_6 \to 0$ so that in the two-dimensional regime we have 
\beq F_{c}^{(2)}(x \to \infty, y \to 0) = 0 \eeq 
and the symmetric  $F_{c}^{(2)}(x \to 0, y \to \infty) = 0$.
Now we are ready to use symbol technology and solve (\ref{FcDiff}). As first step, we evaluate the symbol of $F^{(1)}_c$,
\begin{eqnarray}
\mathcal{S}\left(F^{(1)}_c\right)&=&\frac{y}{x-y}(- [1+x,y]+ [1+x,1+y]- [y,1+x]+ [1+y,1+x]) \nonumber \\
&+& \frac{x}{x-y} \left(- [x,1+x]+ [x,1+y]+ [1+x,x]- [1+x,1+y]+ [y,1+y] \right. \nonumber \\
&& \left.  \qquad \qquad +  [1+y,x]- [1+y,1+x]- [1+y,y]\right)
\end{eqnarray}
where every square bracket is composed by two letters, being the symbol of a transcendentality-two function (we use the notation of appendix \ref{symbols}). The next step is to integrate the symbol using the differential equation and then add terms given by the integrability condition (\ref{integrability}), which assures that the symbol actually corresponds to a function. To the result, we can still add any single-variable symbol $\mathcal{S}(g(x))$ which is invisible for the differential equation and is fixed by imposing the  boundary conditions on the symbol. We finally find that 
{\small{
\begin{eqnarray}
\mathcal{S}(F^{(2)}_c)&=&  [x,1+x,x,1+x]- [x,1+x,x-y,1+x]+ [x,1+x,x-y,1+y]- [x,1+x,y,1+y] \nonumber \\
&+&  [x,1+x,1+y,1+x]- [x,y,1+x,1+y]- [x,y,1+y,1+x]+ [x,1+y,1+x,1+y] \nonumber \\
&+& [x,1+y,x-y,1+x]- [x,1+y,x-y,1+y]+ [x,1+y,y,1+y]- [1+x,x,x,1+x] \nonumber \\
&+& [1+x,x,x-y,1+x]- [1+x,x,x-y,1+y]+ [1+x,x,y,1+y]- [1+x,x,1+y,1+x] \nonumber \\
&+& [1+x,y,x,1+x]- [1+x,y,x-y,1+x]+ [1+x,y,x-y,1+y]+ [1+x,y,1+y,1+x] \nonumber \\
&-& [1+x,1+y,x,1+x]- [1+x,1+y,y,1+y] + (x \leftrightarrow y) \, .
\end{eqnarray}
}}
This symbol corresponds to the one found in \cite{Alday:2010jz} when appropriate differential operator and change of variables are taken into account.
One can now express the symbol in term of functions. To this extent, we have considered the letters $\{x, 1+x, y, 1+y, x-y\}$ and their combinations as arguments of degree four classical harmonic  polylogarithms. Nevertheless, since the constraint (\ref{4degreesymm}) is not satisfied by $\mathcal{S}(F^{(2)}_c)$, the expression cannot be written in terms of classical polylogarithms only. Therefore additional functions must be added, and we find that in case $(c)$ only the following contribute
\begin{eqnarray}
\cW_1 &=& \int_0^x \frac{H_{011}(-z) - H_{011}(-y)}{z-y} dz   \, ,\\
\cW_2 &=& \int_0^y \frac{H_{011}(-z) - H_{011}(-x)}{z-x} dz 
\end{eqnarray}
and appear in the final expression of $F^{(2)}_c$.
Imposing the validity of the differential equation and the boundary conditions implies that the terms to be subtracted are 
\beq
\BB_c = \frac{\pi^4}{60} + 2 \log(1+x) \zeta_3  +\frac{\pi ^2}{6}  \left(\log x \log(1+x) - \frac{1}{2} \log(1+x)^2+ \Li_2(-x)\right) + (x \rightarrow y)
\eeq
which are indeed lower-degree functions multiplied by constants of the right degree. The final expression for the two-loop integral is
\beq
F^{(2)}_c(x,y) = -2 \cW_1 -2 \cW_2 -\BB_c + {\rm classical \, polylogs} \, .
\eeq
The classical-polylogarithm part is lengthy and explicitly showed in appendix \ref{FcHPL}. It deserves  further investigation to be simplified.

 \begin{figure}[t]
\psfrag{dots}[cc][cc]{$\ldots$}
 \centerline{
 {\epsfxsize6cm  \epsfbox{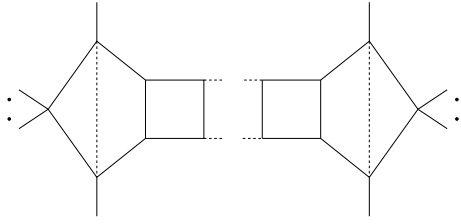}}
} 
 \caption{\small Double pentaladder  integrals which obey the same iterative differential equations.} \label{ladder}
\end{figure}
To conclude this section, we want to remark again that the differential equations appearing in this and in the other cases are valid at all-loop order
\begin{equation}
\OO_{24} \OO_{46} \frac{1}{N_{Lc}} F_{c}^{(L)} = \frac{1}{(3456)} \frac{1}{N_{(L-1)c}} F_c^{(L-1)}\, ,
\end{equation}
where with $ F_{c}^{(L)}$ we mean the double pentaladder integrals of Fig. \ref{ladder}, where the pentagon subintegrals correspond case by case to the ones at two loops. which are present in loop amplitudes.
This iterative structure, relating higher-loop orders to one-loop hexagons, is very powerful since it allows to obtain analytical results from one-loop information. 

\subsection{Double-pentagon integrals $F^{(2)}_{a}$ and $F^{(2)}_{b}$}
\label{othertop}

We now consider the other two topologies appearing in the two-loop eight-point  MHV amplitude. 
The case $F_{a}^{(2)}$ is the simplest, as it turns out to  depend on one cross-ratio only when projected to the two-dimensional subspace. 
Considering the massive pentagon, we have in four dimensions
\begin{equation}\label{diff-massive-a}
\OO_{24} \OO_{86} \frac{1}{(3782)}  F_{a}^{(2)} = -\frac{1}{(3467)} F_{a}^{(1)}\,,
\end{equation}
where $F_{a}^{(1)}$ is the one-loop hexagon integral  with one massive and one massless corner as in Fig. \ref{fig-oneloop}(a).
The reduction to two-dimensional kinematics leaves only one independent cross-ratio and reduces the differential equation to
\begin{eqnarray}
y \partial_y (1+y) \partial_y F_{a}^{(2)}(y) = F_{a}^{(1)}(y) \, ,
\end{eqnarray}
where the boundary condition is $F_{a}^{(2)}(y\rightarrow\infty)=0$. The inhomogeneous term is known \cite{ArkaniHamed:2010gh} and in two dimensions becomes
\begin{eqnarray}
F^{(1)}_{a}(y )  &=& - \log \frac{1}{1+y} \log(\frac{y}{1+y}) - 2 \Li_2 (\frac{1}{1+y}) \, .
\end{eqnarray}
Actually in this particular case it would be easily possible to  directly  write the solution in terms of harmonic polylogarithms. If we nevertheless make use of symbology, starting from the symbol  for $F^{(1)}_{a}(y )$,
\beq
{\mathcal S}(F_{a}^{(1)}) = - \left[\frac{1}{(1+y)} , \frac{y}{(1+y)} \right]+ \left[\frac{y}{(1+y)} , \frac{1}{(1+y)}\right]\, ,
\eeq
and given the dependence of the integrals from one cross-ratio only, we can directly write the result coming from the differential equation at $L$-th loop order
\begin{eqnarray}
{\cal S}(F_{a}^{(L)}) &=& [{\cal S}(F_{a}^{(1)}) \underbrace{, \frac{1}{y}  , \frac{1}{(1+y)} , \ldots , \frac{1}{y}  , \frac{1}{(1+y)}}_{2(L-1) - {\rm fold}\;{\rm tensor}\;{\rm product}} ] \,.
\end{eqnarray}
In terms of functions, the two-loop expression for the case $(a)$ is easily written as
\begin{eqnarray}
F_{a}^{(2)} &=& -\text{Li}^2_2(-y)-2 \text{Li}_4(-y)-2 \text{Li}_4\left(\frac{1}{y+1}\right)-2 \text{Li}_4\left(\frac{y}{y+1}\right) \nn \\
&& -\frac{1}{3}  \pi ^2 \left(\text{Li}_2(-y)-\frac{1}{2} \log ^2(y+1)+\log (y) \log (y+1)\right) \nn \\
&& -\text{Li}_2(-y) \log (y+1) \log (y)+2 \text{Li}_3(-y) \log (y)+2 \text{Li}_3\left(\frac{y}{y+1}\right) \log (y) \nn \\
&& -\text{Li}_3(-y) \log (y+1)-2 \zeta (3) \log(y+1)-\frac{1}{4} \log ^4(y)-\frac{1}{12} \log ^4(y+1) \nn \\
&& -\frac{1}{12} \log ^4\left(\frac{y+1}{y}\right)+\frac{2}{3} \log (y+1) \log^3(y)-\frac{1}{3} \log ^3\left(\frac{y+1}{y}\right) \log (y)\nn \\
&& -\frac{1}{2} \log ^2(y+1) \log ^2(y)-\frac{2 \pi ^4}{45} \, .
\end{eqnarray}

The case $F_{b}^{(2)}$ is more similar to case (c) and the independence on the dual points $x_{1}$ and $x_{5}$ reduces the number of cross-ratios.
Let us consider again the differential operators adapted to the massive corners as in Section \ref{intro}, \emph{i.e}
\begin{equation}\label{diff-massive-b}
\OO_{24} \OO_{75} \frac{1}{(3672)}  F_{b}^{(2)} = -\frac{1}{(3456)} F_{b}^{(1)}\,.
\end{equation}
After utilization of the chain rule and the verification of the compatibility between the differential operator and the  two-dimensional kinematics, we can write:
\begin{eqnarray}
 x \partial_{x} y \partial_{y}  F_{b}^{(2)}(x,y) = F_{b}^{(1)}(x,y) \,,
\end{eqnarray}
where the hexagon diagram is given by \cite{ArkaniHamed:2010gh}
\begin{eqnarray}
F_{b}^{(1)}(x,y) &=&\frac{x }{(y - x)}  \left[2 \, \Li_2(-x) + 2 \zeta_2 - \log x \log y + 
      \log( x y ) \log(1 + x)\right]  + (y \leftrightarrow x) \, .
\label{Fb1}
\end{eqnarray}
As discussed in section \ref{boundary}, the double soft limit $p_5, p_6 \rightarrow 0 $ fixes the boundary conditions to 
\beq F_{b}^{(2)}(x\rightarrow 0,y\rightarrow\infty) = 0 \eeq
and the symmetric one $F_{b}^{(2)}(x\rightarrow\infty,y\rightarrow 0)=0$. Using the differential equation and the integrability condition as before, we arrive at 
\begin{eqnarray}
\mathcal{S}(F_b^{(2)}) &=&  [x,x,y,y]-  [x,x,1+y,y]+  [x,1+x,x-y,x]-  [x,1+x,x-y,y] \nonumber \\
&-&  [x,1+x,y,x]+  [x,1+x,1+y,y]+  [x,y,x,y]-  [x,y,1+x,x] \nonumber \\
&+&  [x,y,y,x]-  [x,y,1+y,y]-  [x,1+y,x,y]+  [x,1+y,1+x,x] \nonumber \\
&-&  [x,1+y,x-y,x]+  [x,1+y,x-y,y]-  [1+x,x,x-y,x]+  [1+x,x,x-y,y] \nonumber \\
&+&  [1+x,x,y,x]-  [1+x,x,y,y]+  [1+x,y,x-y,x]-  [1+x,y,x-y,y] \nonumber \\
&-&  [1+x,y,y,x]+  [1+x,y,1+y,y] + (x\leftrightarrow y) \, .
\end{eqnarray}
From the failure of the constraint (\ref{4degreesymm}), as in case $(c)$, the final result will contain functions different from classical polylogarithms. For this diagram two of such terms are necessary:
\begin{eqnarray}
\cW_3 &=& \int_0^x \frac{H_{001}(-z) - H_{001}(-y)}{z-y} dz   \\ \,
\cW_4 &=& \int_0^y \frac{H_{001}(-z) - H_{001}(-x)}{z-x} dz  \, 
\end{eqnarray}
and the boundary conditions imply the subtraction of the following terms 
\beq
\BB_b = \frac{\pi ^4}{45}+\frac{\pi^2}{12}  \log^2 x -\frac{\pi ^2 }{4} \log^2(1+x)+4  \zeta_3  \log x + 2  \zeta_3  \log(1+x) +\frac{\pi ^2}{3} \log x \log y + (x \rightarrow y) \, ,
\eeq
The final result will therefore be
\beq
F_b^{(2)}(x,y) = - 2 \cW_3 - 2 \cW_4  -\BB_b + {\rm classical \, polylogs} 
\eeq
where the classical-polylogarithm terms are presented in appendix \ref{FbHPL}.

As for case $(c)$, we can extend these differential equations to the ladder diagrams depicted in Fig. \ref{ladder}.

\subsection{Two-loop result}
\label{twoloopcomplete}
We combine now the previous results found for the finite diagrams appearing in the two-dimensional two-loop MHV scattering amplitude. To do so, we must consider the sum over cyclic permutations of each diagram. 
This is taken into account by evaluating for the three different cases:
\begin{eqnarray}
F_{a}^{(2)} &=& F_{a}^{(2)}(x) + F_{a}^{(2)}(y) + F_{a}^{(2)}\left(\frac{1}{x}\right) + F_{a}^{(2)}\left(\frac{1}{y}\right) \\
F_{b,c}^{(2)} &=& F_{b,c}^{(2)}(x,y) + F_{b,c}^{(2)}\left(y,\frac{1}{x}\right) + F_{b,c}^{(2)}\left(\frac{1}{x},\frac{1}{y}\right) + F_{b,c}^{(2)}\left(\frac{1}{y},x\right)
\end{eqnarray}
and considering twice the diagrams $(a)$ and $(c)$.
The symbol is then given  by
\begin{eqnarray}
\mathcal{S}(F^{(2)}) &=&  [x,x,y,y]-2  [x,1+x,x,y]+4  [x,1+x,x,1+y]-2  [x,1+x,y,x] \nonumber \\ 
&+& 4  [x,1+x,1+y,x]+ [x,y,x,y]-2  [x,y,1+x,x]  + [x,y,y,x]   \nonumber \\  
&-& 2  [x,y,1+y,y]+4  [x,1+y,1+x,x]+2  [x,1+y,y,y] + 2  [1+x,x,x,y]   \nonumber \\ 
&-& 4  [1+x,x,x,1+y]+2  [1+x,x,y,x]-4  [1+x,x,1+y,x]+2  [1+x,y,x,x]  \nonumber \\ 
&+& 4  [1+x,y,1+y,y]-4  [1+x,1+y,x,x]-4  [1+x,1+y,y,y]+ [y,x,x,y]  \nonumber \\ 
&-& 2  [y,x,1+x,x]+ [y,x,y,x]-2  [y,x,1+y,y]+2  [y,1+x,x,x]  \nonumber \\ 
&+& 4  [y,1+x,1+y,y]+ [y,y,x,x]-2  [y,1+y,x,y]+4  [y,1+y,1+x,y]  \nonumber \\ 
&-& 2  [y,1+y,y,x]+4  [y,1+y,y,1+x]+4  [1+y,x,1+x,x]+2  [1+y,x,y,y]   \nonumber \\ 
&-& 4  [1+y,1+x,x,x]-4  [1+y,1+x,y,y]+2  [1+y,y,x,y]-4  [1+y,y,1+x,y] \nonumber \\  
&+& 2  [1+y,y,y,x]-4  [1+y,y,y,1+x] \, .
\end{eqnarray}
We want to remark here that all the terms with the $(x-y)$ letter have disappeared. This was indeed noticed already through the differential equations of \cite{Alday:2010jz}.
The expression for the highest-degree terms of the final function drastically simplifies:
\begin{eqnarray}
F^{(2)}_{\rm highest} &=& 2  \text{Li}_2(-x) \log (x) \log (y)-4 \text{Li}_2(-y) \log (1+x) \log (y)+2 \text{Li}_2(-y) \log (x) \log (y)  \nonumber \\ 
&-& 6  \text{Li}_3(-x) \log (y)-4 \text{Li}_2(-x) \log (x) \log (1+y)+12 \text{Li}_3(-x) \log (1+y)  \nonumber \\ 
&+& 12  \text{Li}_3(-y) \log (1+x)-6 \text{Li}_3(-y) \log (x)+\frac{1}{4} \log ^2(x) \log ^2(y) \, .
\end{eqnarray}
We have checked that taking the $x$ and $y$ derivatives of $F^{(2)}_{\rm highest} $, we find the same function of \cite{Alday:2010jz}.
One important point that we want to stress here is that the functions which were not expressible in terms of classical polylogarithms have now disappeared: if we indeed use the relation (\ref{4degreesymm}) for the symbol of $F^{(2)}$ we correctly find that it vanishes.

\subsection{Two-loop $n$-point and $L$-loop ladder integrals in two dimensions}

 \begin{figure}[t]
\psfrag{dots}[cc][cc]{$\ldots$}
 \centerline{
 {\epsfxsize4cm  \epsfbox{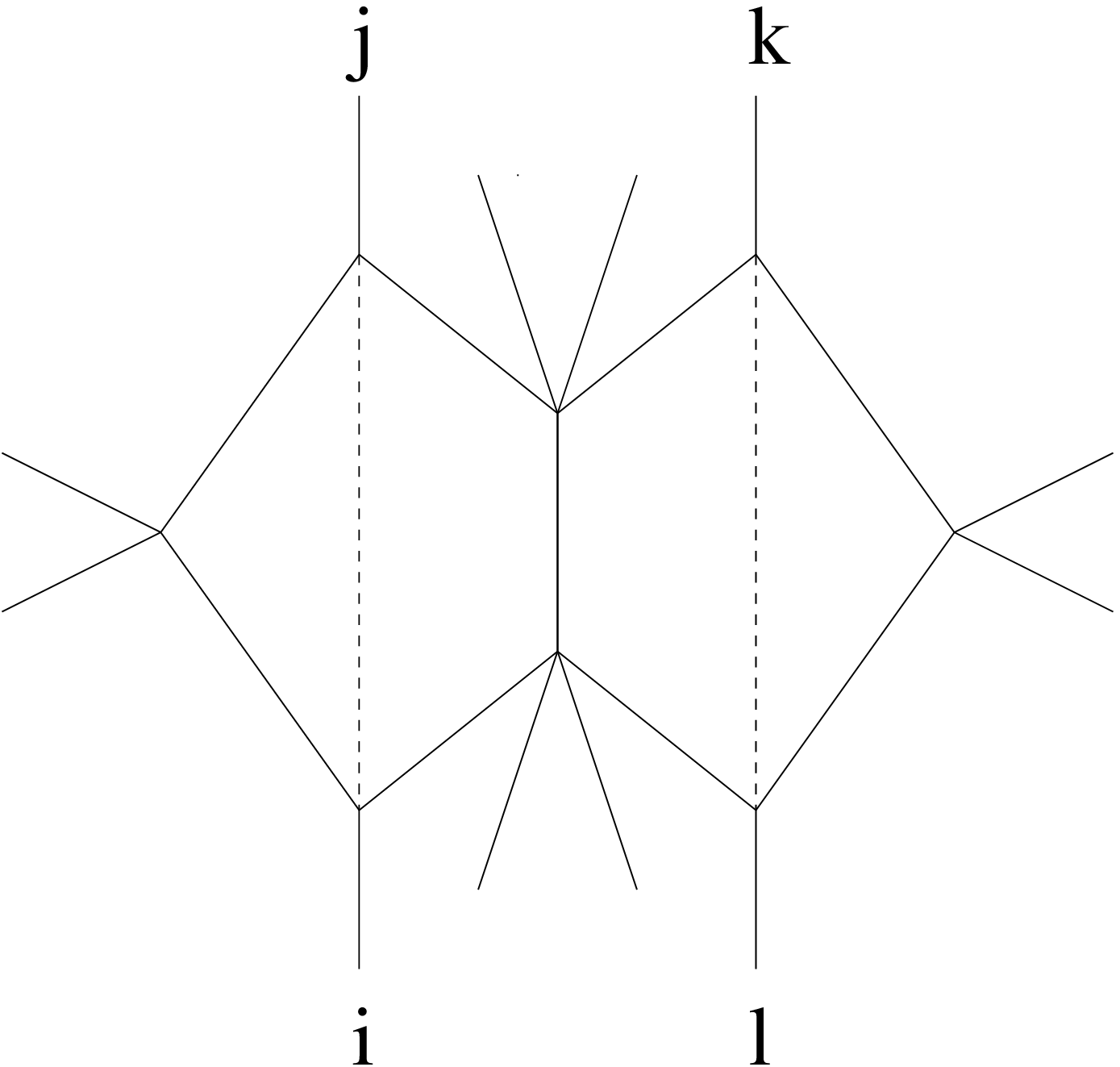}}
\qquad\quad  {\epsfxsize4cm  \epsfbox{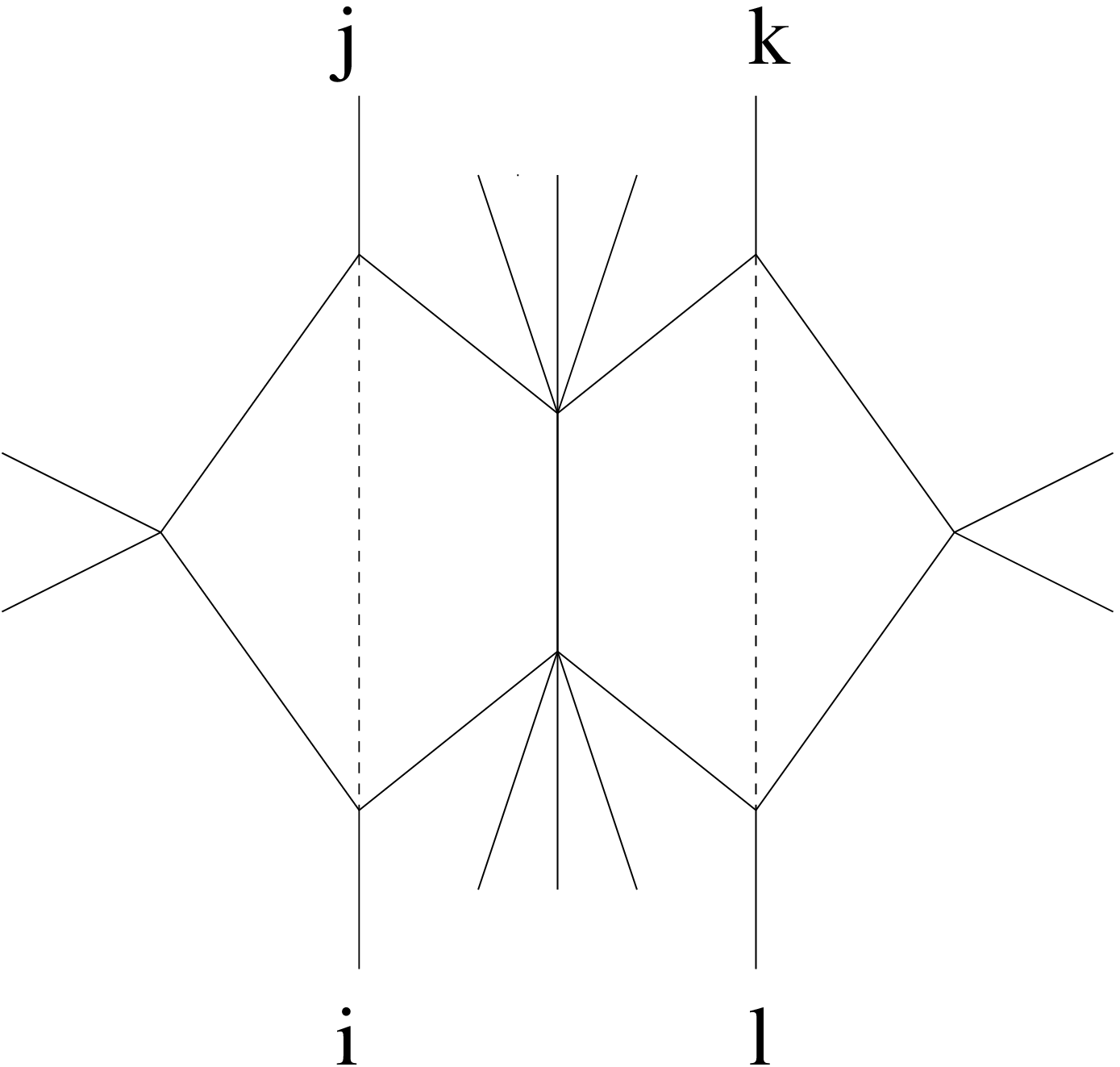}}
\qquad\quad  {\epsfxsize4cm  \epsfbox{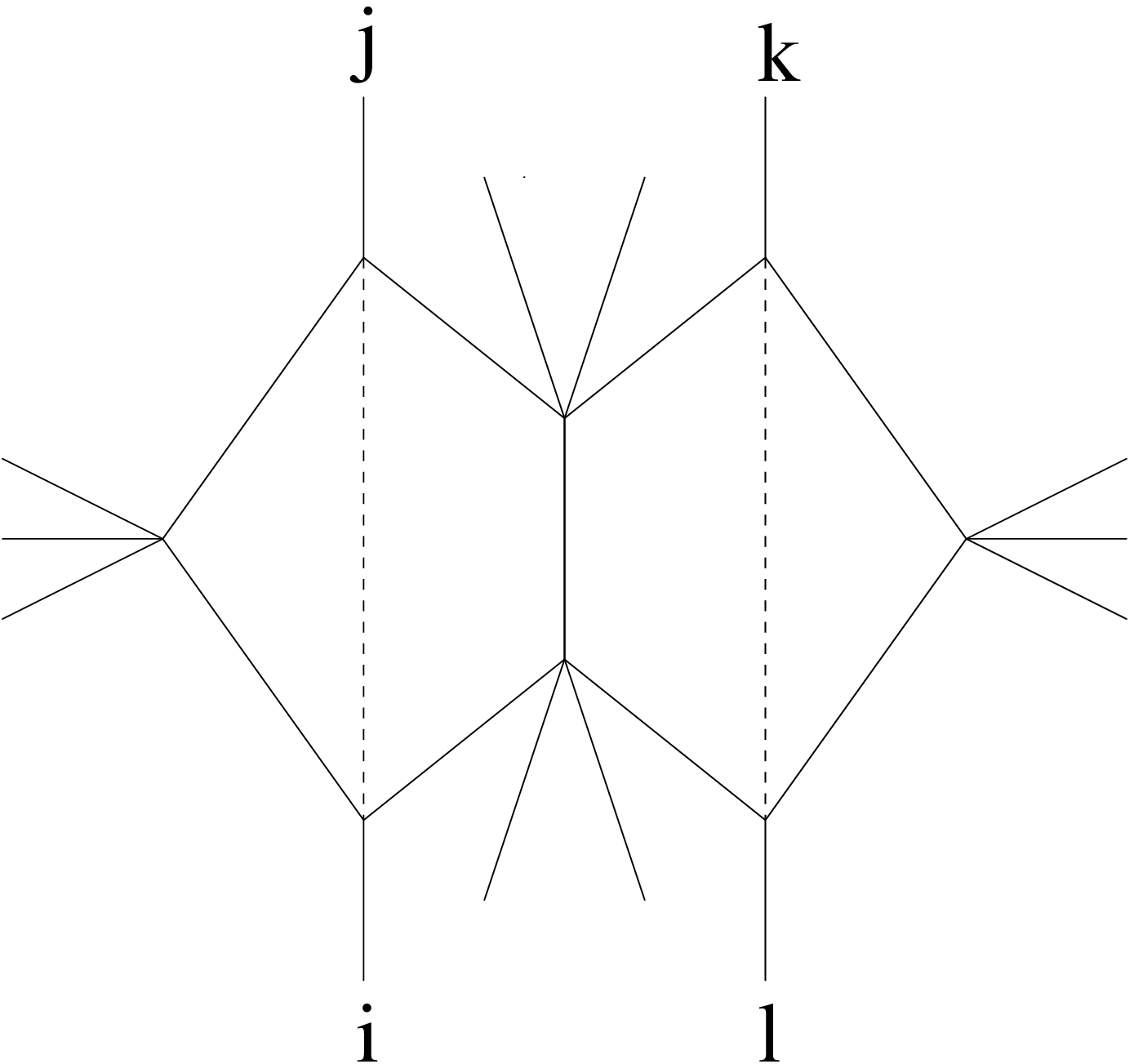}}
} 
 \caption{\small Two-loop n-point integrals which survive in the two-dimensional kinematics.} 
\label{2dnpt}
\end{figure}

We want to consider now particular diagrams appearing in scattering amplitudes with high number of legs and at higher loop, to see which topologies survive to the two-dimensional reduction. 

We start by looking at double-pentagon integrals with arbitrary number of legs in four dimensions.
To this extent, we use the Feynman parametrization,
\beq
\frac{1}{A_1^{\lambda_1}... A_n^{\lambda_n}} = \frac{\Gamma(\lambda_1+...+\lambda_n)}{\Gamma(\lambda_1)...\Gamma(\lambda_n)} \int_0^{\infty} d\alpha_1...d\alpha_n \delta(\sum\alpha_i -1) \frac{\alpha_1^{\lambda_1-1}...\alpha_n^{\lambda_n-1}}{(\alpha_1 A_1 +...+\alpha_n A_n)^{\lambda_1+...+\lambda_n}}
\eeq 
to rewrite the general formula for double pentagons
\beq
\label{general}
\FF^{(2)}_n = N \int \frac{d^4Z_{AB}}{i \pi^2} \frac{d^4Z_{CD}}{i \pi^2} \frac{(A B i j)(C D k l)}{(AB \,a-1\,a)...(ABCD)...(CD\,b-1\,b)} \, ,
\eeq
where $i, j, k$ and $l$ denote the legs of the tensor numerators in the pentagons, in terms of the Feynman parameters $\alpha_i$ and $\beta_i$
\beq
\label{feynman}
\FF^{(2)}_n \propto  \int_0^{\infty} \Pi_i d\alpha_i d\beta_i \delta(\sum\alpha_i -1) \delta(\sum\beta_i -1) \frac{f(\alpha_i, \beta_i) }{\MM}  \left(6 \frac{(i j Y) (k l Y)}{(Y Y)^4}-\frac{(i j k l) }{(Y Y)^3}\right) \, .
\eeq
In going from (\ref{general}) to (\ref{feynman}) we have performed several steps. At first, one has to Feynman parametrize one of the two loops, say the $(AB)$ one,  and then  integrate over the $d^4 Z_{AB}$ following the techniques of \cite{Drummond:2010mb}. After the integration of a specific Feynman parameter, the same steps for the second loop $(CD)$ leads to  (\ref{feynman}). For what we want to argue, we do not need to specify the function $f(\alpha, \beta)$, neither the $\MM$ denominator, which is just a combination of four-brackets of the type $(Z_{a-1} \, Z_{a} \,Z_{b-1} \, Z_b )$.
$Y$ instead is a particular bitwistor: it is indeed a linear combination of bitwistors of the form $(Z_{m-1}, Z_m)$ with coefficients the Feynman parameters. This particular form is given by the parametrization, which combines indeed the propagators in one denominator.
The interesting term for what we are going to discuss, is the combination in the bracket
\beq
\label{bracket}
6 \frac{(i j Y) (k l Y)}{(Y Y)^4}-\frac{(i j k l) }{(Y Y)^3} \, .
\eeq
The reduction to two dimensions makes  indeed  vanish some or even all the terms in (\ref{bracket}) in particular configurations. Given the special form of $Y$, the first term will survive only when $(i j)$ and $(k l)$ $ \sim ({{\rm even ~odd}})$. Concerning the second term, many more combinations will survive for $(i j k l)$, provided that two of them are odd and two even.
Requiring the number of legs to be even, as we are in two dimensions, the only topologies which survive are depicted in Fig. \ref{2dnpt}, where with two (three) legs we mean an even (odd) number of them.

 \begin{figure}[t]
\psfrag{dots}[cc][cc]{$\ldots$}
 \centerline{
 {\epsfxsize5cm  \epsfbox{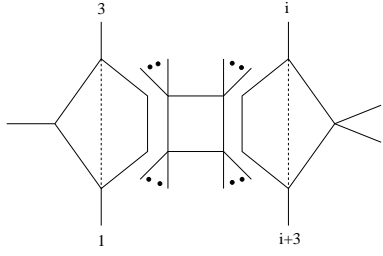}}
} 
\caption{\small The double pentaladder diagrams which simplify in two dimensions.} \label{fig-vanishing}
\end{figure}

We want now to comment on the  reduction to two dimensions for the double pentaladder integrals of Fig.   \ref{fig-vanishing}, where any number of legs is inserted between the pentagons and the boxes. When written in the Feynman parametrization they will contain again terms of the type $(1 3 i i+3)$ and $(1 3 Y)$, where $Y$ is  a bitwistor formed by some linear combination of $(Z_{m-1}, Z_m)$ as before.
 Therefore, due to the particular structure of these diagrams, it actually happens that they are subject to strong simplifications when projected to two dimensions. 
 Moreover, using the differential operators we have worked out and in particular the ones similar to the case $(b)$, it can be showed that at three loop they are reduced to two-loop diagrams which vanish in two dimensions, given the comments above (see  Fig. \ref{reduction2} for an example). This is a very powerful constraint for the function, as it fixes the inhomogeneous term of the differential equation to be zero.
 \beq
\label{reduction2}
\mathcal{D}^{(2)}~
{
\parbox[c]{40mm}{\includegraphics[height = 3cm] {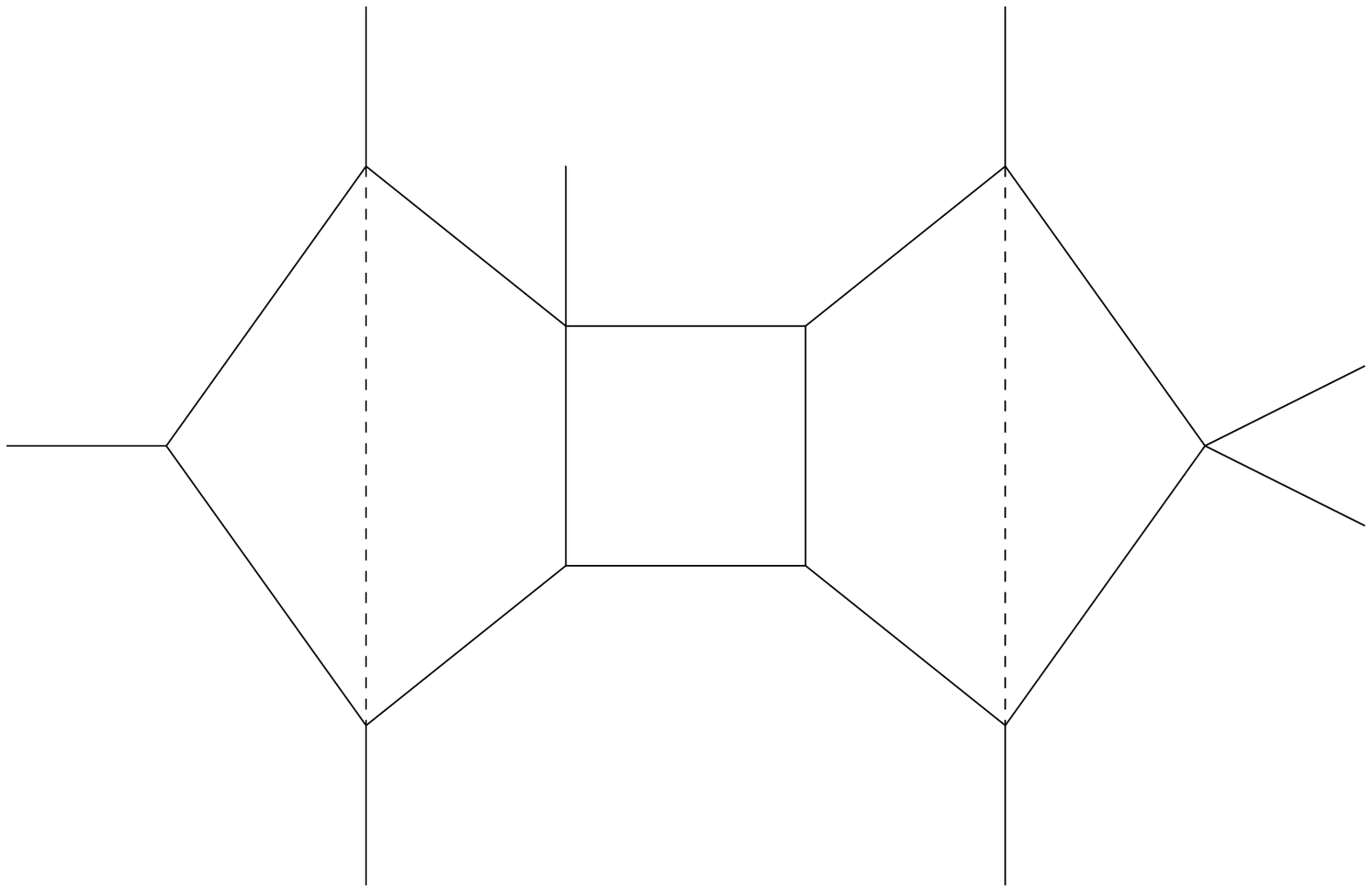}}
}
\qquad \propto \,
{
\parbox[c]{30mm}{\includegraphics[height = 3cm] {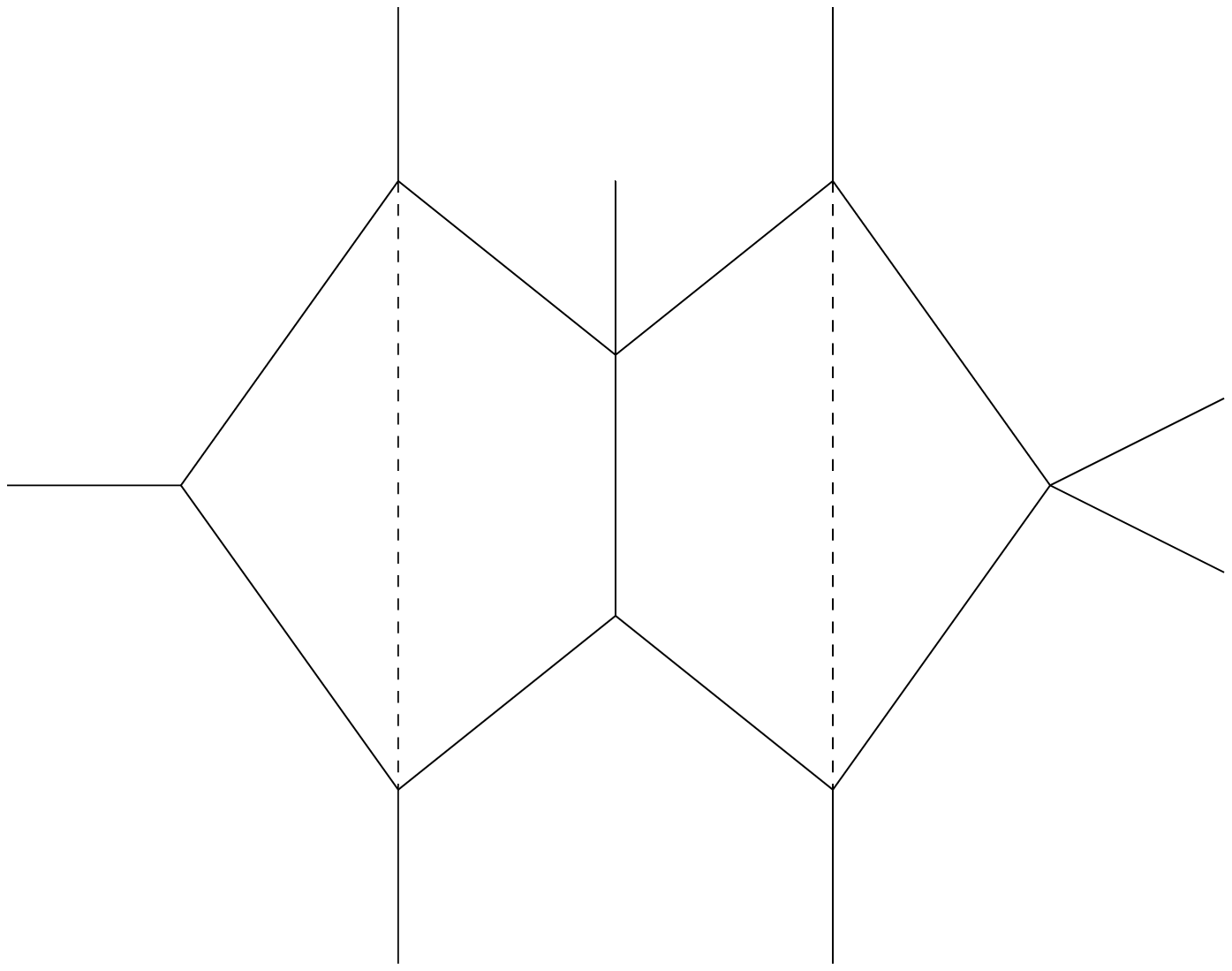}}
}
\qquad =   0 \, ({\rm two-dim})
\eeq

Therefore, given the differential operators and the two-dimensional kinematics, the analytic results for the topologies present in $L$-loop $n$-point scattering amplitudes  strongly simplify.

\section*{Acknowledgments}
I would like to thank J. Drummond and J. Henn for collaboration at different stages of this project. This work was supported by the Volkswagen-Foundation.

\appendix

\section{Review on symbology}
\label{symbols}
In this appendix we want to recall the notion of symbol and its properties. 
A transcendental function of degree $k$ is defined iteratively as the linear combination with rational coefficients of $k$ iterated integrals:
\begin{eqnarray}
f_k &=& \int_a^b d\log R_1 \otimes ... \otimes d\log R_k \\
&=& \int_a^b d\log R_k(t) \left[\int_0^t d\log R_1 \otimes ... \otimes d\log R_{k-1}\right] \, , 
\end{eqnarray}
where $R_i$ are rational functions. For instance the trilogarithm $\Li_3(z)$  is given by
\beq
\Li_3(z) = - \int_0^z d\log t_1 \int_0^{t_1} d\log t_2 \int_0^{t_2} d\log (1-t_3) \, .
\eeq
The symbol of the pure function $f_k$ is defined as
\beq
\mathcal{S}(f_k) = R_1 \otimes ... \otimes R_{k} \equiv \left[R_1,..., R_k\right] \, ,
\eeq
where for sake of simplicity throughout the paper we use a square-bracket notation.  For instance the symbol of $\Li_3(z)$ is
\beq
\mathcal{S}(\Li_3(z)) = - [1-t, t, t] \, .
\eeq
The  symbol of a function makes clear the locations of the discontinuities of the function.
It satisfies, besides the shuffle algebra, many properties which are useful in simplifying expressions. For instance, for given rational functions $A, B$ and  $ \frac{a_i a_j}{a_k}$ we have 
\beq
\left[A, \frac{a_i a_j}{a_k}, B\right] = \left[A, a_i, B\right] + \left[A, a_j, B\right]-\left[A, a_k, B\right] \, ,
\eeq
while if $c$ is a constant different from zero, the symbol is invisible to it:
\beq
\left[A, c~ a_i, B\right] = \left[A, a_i, B\right] \, .
\eeq
The derivative of a symbol is instead performed acting on its last entry:
\beq
\frac{\partial}{\partial x} [a_1,...,a_n] = [a_1,...,a_{n-1}] \times  \frac{\partial}{\partial x} \log a_n \, .
\eeq
The symbol loses some information about the function. Besides missing the knowledge of  which logarithmic branch the integrand 
of an iterated integral is on,  at a given transcendental degree it does not detect functions which are transcendental constants times pure functions of lower degree.
Let us consider for instance the following identity
\beq
\Li_2(-z) + \Li_2\left(-\frac{1}{z}\right) = -\frac{1}{2} \log^2(z)-\frac{\pi^2}{6} \, ,
\eeq
and compute the symbol of the left-hand side
\beq
\mathcal{S}(\Li_2(-z)) + \mathcal{S}\left(\Li_2\left(-\frac{1}{z}\right)\right) = - [z,z] \, .
\eeq
Recalling the symbol of the $n$-th power of the logarithm
\beq
\mathcal{S}\left(\log^n(z)\right)  = n! \, \underbrace{[z,..., z]}_{n-times} \, ,
\eeq
we see that the $\frac{\pi^2}{6}$ term is missed. \\
\indent
Due to the fact that $d^2 f_k = 0$, the letters of its symbol $[R_1,...,R_i, R_{i+1},...R_k]$ are not all independent, but constrained by the integrability condition
\beq
\label{integrability}
d\log R_i \wedge d\log R_{i+1} [R_1,...,R_{i-1}, R_{i+2},...R_k] = 0 \, .
\eeq 
This means that, if to a pure function always corresponds a symbol, the other way round is not always true. For instance $[x, y]$ is not the symbol of a function, since it does not satisfy (\ref{integrability}), but $\left([x, y] + [y, x]\right)$ is. 
There is a further constraint that we want to mention here, as it is used in the main text.  For degree $k \leqslant 4$, Goncharov conjectured that symbols obeying 
\beq
\label{4degreesymm}
\mathcal{S}^{(k)}_{abcd} - \mathcal{S}^{(k)}_{bacd} - \mathcal{S}^{(k)}_{abdc} + \mathcal{S}^{(k)}_{badc} - \left(a \leftrightarrow c, b\leftrightarrow d\right) = 0
\eeq
can be obtained from a function involving classical polylogarithms only.

\section{Explicit expression for the classical HPLs in $F_b^{(2)}$}
\label{FbHPL}
In the following, we present the expression given by the symbol we have found for $F_b^{(2)}$ when written in terms of HPLs. We recall that, in order to get the complete answer, two functions which are not classical HPLs and lower-degree functions multiplied by constants must be added.
{\footnotesize
\begin{eqnarray}
F_{b,class}^{(2)}&=& \frac{\log ^4(x)}{12}-\frac{1}{6} \log (y) \log ^3(x)+\frac{1}{4} \log ^2(x+1) \log ^2(x)+\frac{1}{4} \log ^2(y) \log ^2(x)-\frac{1}{4} \log ^2(y+1) \log ^2(x) \nonumber \\
&+&\frac{1}{2} \text{Li}_2(-x) \log ^2(x)-\frac{1}{2} \text{Li}_2\left(1-\frac{x+1}{y+1}\right) \log ^2(x)  +\frac{1}{12} \log ^3(x+1) \log (x)+\frac{1}{6} \log ^3(y) \log (x) \nonumber \\
&+&\frac{1}{12} \log ^3(y+1) \log (x)-\frac{1}{4} \log (x+1) \log ^2(y+1) \log (x)+\frac{1}{4} \log ^2(x+1) \log (y+1) \log (x) \nonumber \\
&-&\log (x+1) \text{Li}_2(-x) \log (x)+\log (y) \text{Li}_2(-x) \log (x)-\frac{1}{2} \log (y+1) \text{Li}_2(-x) \log (x) \nonumber \\
&+&\frac{1}{2} \log (x+1) \text{Li}_2\left(1-\frac{x}{y}\right) \log (x)-\frac{1}{2} \log (y+1) \text{Li}_2\left(1-\frac{x}{y}\right) \log (x)-\frac{1}{2} \log (x+1) \text{Li}_2(-y) \log (x) \nonumber \\
&+&\log (y) \text{Li}_2(-y) \log (x)+\text{Li}_3(-x) \log (x)-3 \text{Li}_3(-y) \log (x)-\frac{5}{2} H_{1,0,1}(-x) \log (x) \nonumber \\
&+&\frac{1}{2} H_{1,0,1}\left(\frac{x-y}{x}\right) \log (x)-\frac{1}{2} H_{1,0,1}(-y) \log (x)-\frac{1}{2} H_{1,0,1}\left(\frac{y-x}{(x+1) y}\right) \log (x) \nonumber \\
&+&\frac{1}{2} H_{1,0,1}\left(1-\frac{x+1}{y+1}\right) \log (x)-\frac{1}{24} \log ^4(x+1)-\frac{\log ^4(y)}{12}-\frac{1}{24} \log ^4(y+1)+\frac{1}{4} \log (y) \log ^3(y+1) \nonumber \\
&-&\frac{1}{4} \log ^2(x+1) \log ^2(y)+\frac{1}{4} \log ^2(y) \log ^2(y+1)-\frac{1}{4} \log (x+1) \log (y) \log ^2(y+1)-\frac{3}{2} \text{Li}^2_2(-x){} \nonumber \\
&-&\frac{3}{2} \text{Li}^2_2(-y){}-\frac{1}{12} \log ^3(x+1) \log (y)+\frac{1}{4} \log ^2(x+1) \log (y) \log (y+1)+\frac{1}{2} \log ^2(x+1) \text{Li}_2(-x) \nonumber \\
&-&\frac{1}{2} \log (y) \log (y+1) \text{Li}_2(-x)+\frac{1}{2} \log (x+1) \log (y) \text{Li}_2\left(1-\frac{x}{y}\right)-\frac{1}{2} \log (y) \log (y+1) \text{Li}_2\left(1-\frac{x}{y}\right) \nonumber \\
&+&\frac{1}{2} \log ^2(y) \text{Li}_2(-y)+\frac{1}{2} \log ^2(y+1) \text{Li}_2(-y)-\frac{1}{2} \log (x+1) \log (y) \text{Li}_2(-y)-\log (y) \log (y+1) \text{Li}_2(-y) \nonumber \\
&-&\text{Li}_2(-x) \text{Li}_2(-y)+\frac{1}{2} \log ^2(y) \text{Li}_2\left(1-\frac{x+1}{y+1}\right)-2 \log (x+1) \text{Li}_3(-x)-3 \log (y) \text{Li}_3(-x) \nonumber \\
&+&\log (y) \text{Li}_3(-y)-2 \log (y+1) \text{Li}_3(-y)+\log (x+1) H_{1,0,1}(-x)-\frac{1}{2} \log (y) H_{1,0,1}(-x) \nonumber \\
&+&\frac{1}{2} \log (y) H_{1,0,1}\left(\frac{x-y}{x}\right)-\frac{5}{2} \log (y) H_{1,0,1}(-y)+\log (y+1) H_{1,0,1}(-y)-\frac{1}{2} \log (y) H_{1,0,1}\left(\frac{y-x}{(x+1) y}\right) \nonumber \\
&+&\frac{1}{2} \log (y) H_{1,0,1}\left(1-\frac{x+1}{y+1}\right)+3 H_{0,1,0,1}\left(\frac{1}{x+1}\right)+3 H_{0,1,0,1}\left(\frac{1}{y+1}\right)+2 H_{1,1,0,1}(-x) \nonumber \\
&+&2 H_{1,1,0,1}\left(\frac{1}{x+1}\right)+2 H_{1,1,0,1}(-y)+2 H_{1,1,0,1}\left(\frac{1}{y+1}\right)
\end{eqnarray}
}

\section{Explicit expression for the classical HPLs in $F_c^{(2)}$}
\label{FcHPL}
{\footnotesize
\begin{eqnarray}
F_{c,class}^{(2)}&=& \frac{5}{24} \log ^4(x+1)-\frac{7}{12} \log (x) \log ^3(x+1)+\frac{3}{4} \log (y) \log ^3(x+1)-\log (y+1) \log ^3(x+1) \nonumber \\
&+&\frac{3}{2} \log ^2(y+1) \log ^2(x+1)+\frac{3}{4} \log (x) \log (y+1) \log ^2(x+1)-\frac{3}{4} \log (y) \log (y+1) \log ^2(x+1)   \nonumber \\
&- &\frac{5}{6} \text{Li}_2(-x) \log ^2(x+1)-\frac{3}{2} \text{Li}_2\left(1-\frac{x}{y}\right) \log ^2(x+1)+\frac{3}{2} \text{Li}_2(-y) \log ^2(x+1) +\text{Li}_3(-y) \log (x+1)  \nonumber \\
&+&\text{Li}_2\left(1-\frac{x+1}{y+1}\right) \log ^2(x+1)+\frac{1}{4} \log ^3(x) \log (x+1)-\frac{1}{4} \log ^3(y) \log (x+1) -2 \text{Li}_3\left(\frac{x-y}{x}\right) \log (x+1)  \nonumber \\
&-&\log ^3(y+1) \log (x+1)+\frac{3}{4} \log (x) \log ^2(y) \log (x+1)+\frac{1}{4} \log (x) \log ^2(y+1) \log (x+1)  \nonumber \\
&+&\frac{3}{4} \log (y) \log ^2(y+1) \log (x+1)-\frac{3}{4} \log ^2(x) \log (y) \log (x+1)-\log (x) \log (y) \log (y+1) \log (x+1)  \nonumber \\
&+&\frac{1}{2} \log (x) \text{Li}_2(-x) \log (x+1)-\frac{1}{2} \log (y) \text{Li}_2(-x) \log (x+1)+\frac{5}{2} \log (y+1) \text{Li}_2(-x) \log (x+1)  \nonumber \\
&+&\log (x) \text{Li}_2\left(1-\frac{x}{y}\right) \log (x+1)-\log (y) \text{Li}_2\left(1-\frac{x}{y}\right) \log (x+1)+\log (y+1) \text{Li}_2\left(1-\frac{x}{y}\right) \log (x+1)  \nonumber \\
&-&\frac{3}{2} \log (x) \text{Li}_2(-y) \log (x+1)+\frac{1}{2} \log (y) \text{Li}_2(-y) \log (x+1)-\frac{3}{2} \log (y+1) \text{Li}_2(-y) \log (x+1)   \nonumber \\
&-& \frac{1}{2} \log (x) \text{Li}_2\left(1-\frac{x+1}{y+1}\right) \log (x+1)+\frac{1}{2} \log (y) \text{Li}_2\left(1-\frac{x+1}{y+1}\right) \log (x+1)  \nonumber \\
&-&2 \log (y+1) \text{Li}_2\left(1-\frac{x+1}{y+1}\right) \log (x+1) +2 \log (y+1) \text{Li}_3\left(\frac{y-x}{(x+1) y}\right)+2 \log (y+1) \text{Li}_3\left(1-\frac{x+1}{y+1}\right)  \nonumber \\
&-& 2 \text{Li}_3\left(\frac{y-x}{(x+1) y}\right) \log (x+1)-2 \text{Li}_3\left(1-\frac{x+1}{y+1}\right) \log (x+1)+\frac{5}{6} H_{1,0,1}(-x) \log (x+1)  \nonumber \\
&-&\frac{1}{2} H_{1,0,1}\left(\frac{x-y}{x}\right) \log (x+1)-\frac{1}{2} H_{1,0,1}(-y) \log (x+1)-\frac{1}{2} H_{1,0,1}\left(\frac{y-x}{(x+1) y}\right) \log (x+1)   \nonumber \\
&-&\frac{1}{2} H_{1,0,1}\left(1-\frac{x+1}{y+1}\right) \log (x+1)+\frac{5}{24} \log ^4(y+1)+\frac{1}{12} \log (x) \log ^3(y+1)-\frac{1}{4} \log (y) \log ^3(y+1)   \nonumber \\
&+&\frac{3}{2} \text{Li}^2_2(-x){}+\frac{3}{2} \text{Li}^2_2(-y){}-\frac{1}{4} \log ^3(x) \log (y+1)+\frac{1}{4} \log ^3(y) \log (y+1)-\frac{3}{4} \log (x) \log ^2(y) \log (y+1)   \nonumber \\
&+&\frac{3}{4} \log ^2(x) \log (y) \log (y+1)-\frac{1}{2} \log ^2(y+1) \text{Li}_2(-x)-\frac{1}{2} \log (x) \log (y+1) \text{Li}_2(-x)   \nonumber \\
&-&\frac{1}{2} \log (y) \log (y+1) \text{Li}_2(-x)+\frac{1}{2} \log ^2(y+1) \text{Li}_2\left(1-\frac{x}{y}\right)-\log (x) \log (y+1) \text{Li}_2\left(1-\frac{x}{y}\right)   \nonumber \\
&+&\log (y) \log (y+1) \text{Li}_2\left(1-\frac{x}{y}\right)+\frac{7}{6} \log ^2(y+1) \text{Li}_2(-y)+\frac{1}{2} \log (x) \log (y+1) \text{Li}_2(-y)   \nonumber \\
&-&\frac{1}{2} \log (y) \log (y+1) \text{Li}_2(-y)-2 \text{Li}_2(-x) \text{Li}_2(-y)+\log ^2(y+1) \text{Li}_2\left(1-\frac{x+1}{y+1}\right)  \nonumber \\
&+&\frac{1}{2} \log (x) \log (y+1) \text{Li}_2\left(1-\frac{x+1}{y+1}\right)-\frac{1}{2} \log (y) \log (y+1) \text{Li}_2\left(1-\frac{x+1}{y+1}\right)  \nonumber \\
&-&2 \text{Li}_2(-x) \text{Li}_2\left(1-\frac{x+1}{y+1}\right)+2 \text{Li}_2(-y) \text{Li}_2\left(1-\frac{x+1}{y+1}\right)+\log (y+1) \text{Li}_3(-x) \nonumber \\
&+&2 \log (y+1) \text{Li}_3\left(\frac{x-y}{x}\right) -\log (y) H_{1,0,1}\left(1-\frac{x+1}{y+1}\right)+\frac{1}{2} \log (y+1) H_{1,0,1}\left(1-\frac{x+1}{y+1}\right)  \nonumber 
\end{eqnarray}
\begin{eqnarray}
&+&\log (y) H_{1,0,1}(-x)-\frac{1}{2} \log (y+1) H_{1,0,1}(-x)+\frac{1}{2} \log (y+1) H_{1,0,1}\left(\frac{x-y}{x}\right)+\log (x) H_{1,0,1}(-y)  \nonumber \\
&+&\frac{5}{6} \log (y+1) H_{1,0,1}(-y)+\frac{1}{2} \log (y+1) H_{1,0,1}\left(\frac{y-x}{(x+1) y}\right)-\log (x) H_{1,0,1}\left(1-\frac{x+1}{y+1}\right) \nonumber \\
&+& \frac{4}{3} H_{1,1,0,1}(-y)  + H_{0,1,0,1}\left(\frac{1}{y+1}\right) +\frac{4}{3} H_{1,1,0,1}(-x) +H_{0,1,0,1}\left(\frac{1}{x+1}\right) 
\end{eqnarray}
}

\section{Special choice of twistors for two-dimensional kinematics}
\label{appendix-2dim}

We implement the  two-dimensional kinematics by making this special choice for the momentum twistor variables:
\begin{eqnarray}
\label{fixing}
Z_1=\left( \begin{array}{c}  0 \\  0 \\  i \sqrt{2} \\    - i \sqrt{2} \\  \end{array}  \right),~~~
Z_2=\left(  \begin{array}{c}   -i \sqrt{2} \\  \frac{i}{\sqrt{2}} \\    0 \\  0 \\    \end{array}   \right),~~~
 Z_3=\left(  \begin{array}{c}  0 \\ 0 \\   0 \\  \frac{i}{\sqrt{2}} \\   \end{array}  \right),~~~
    Z_4=\left(  \begin{array}{c}  i \sqrt{2} y \\  \frac{i (1-y)}{\sqrt{2}} \\  0 \\  0 \\  \end{array}  \right) \nonumber \\
      Z_5=\left(  \begin{array}{c}   0 \\   0 \\   i \sqrt{2} x \\   \frac{i (1-x)}{\sqrt{2}} \\  \end{array}  \right),~~~
    Z_6=\left(    \begin{array}{c}  0 \\  \frac{i}{\sqrt{2}} \\   0 \\     0 \\   \end{array}  \right),~~~
       Z_7=\left(  \begin{array}{c}  0 \\  0 \\  - i \sqrt{2} \\   \frac{i}{\sqrt{2}} \\  \end{array}   \right),~~~
     Z_8=\left(  \begin{array}{c}  i \sqrt{2}  \\ - i \sqrt{2}   \\  0 \\   0 \\  \end{array}  \right)
\end{eqnarray}

\section{Definitions of $F_{a,b,c}^{(1)}$ and $F_{a,b,c}^{(2)}$ diagrams}
\label{defloops}
In this appendix we give the definition of two-loop double-pentagon and one-loop hexagon integrals, with the respective normalizations,  which are used throughout the paper:  
\subsection*{Integrals of type (a)}
\begin{equation}
F_{a}^{(2)} = \int \frac{d^4Z_{AB}}{i \pi^2} \frac{d^4Z_{CD}}{i \pi^2} \frac{(2378) (1234) (6781)  (AB82) (CD37) }{(AB12) (AB23)(AB78)(AB81)(ABCD)(CD23)(CD34)(CD67)(CD78)}
\end{equation}
\begin{equation}
F_{a}^{(1)} = \int \frac{d^4Z_{AB}}{i \pi^2}  \frac{(6781)(1234)(AB37)(AB28)}{(AB12) (AB23)(AB34)(AB67)(AB78)(AB81)}
\end{equation}

\subsection*{Integrals of type (b)}
\begin{equation}
F_{b}^{(2)} = \int \frac{d^4Z_{AB}}{i \pi^2} \frac{d^4Z_{CD}}{i \pi^2} \frac{(2367) (1234)(5678) (AB27) (CD36)}{(AB12) (AB23)(AB67)(AB78)(ABCD)(CD23)(CD34)(CD56)(CD67)}
\end{equation}
\begin{equation}
F_{b}^{(1)} = \int \frac{d^4Z_{AB}}{i \pi^2} \frac{ (1234)(5678)(AB36)(AB72)}{(AB12) (AB23)(AB34)(AB56)(AB67)(AB78)}
\end{equation}

\subsection*{Integrals of type (c)}
\begin{equation}
F_{c}^{(2)} = \int \frac{d^4Z_{AB}}{i \pi^2} \frac{d^4Z_{CD}}{i \pi^2} \frac{N_{2c} (AB28) (CD35) }{(AB12) (AB23)(AB78)(AB81)(ABCD)(CD23)(CD34)(CD45)(CD56)}
\end{equation}
where
\begin{equation}
N_{2c} = -(1234)\left[(3781)(5624)+(7812)(3456)\right]
\end{equation}
\begin{equation}
F_{c}^{(1)} = \int \frac{d^4Z_{AB}}{i \pi^2}  \frac{(1234)(6781) (AB28) (AB35) }{(AB12) (AB23) (AB34) (AB56) (AB78)(AB81)}
\end{equation}

\end{document}